\documentclass{aa}  
\usepackage{natbib,twoopt}
\usepackage{impnattypo}
\usepackage[breaklinks=true,colorlinks=true,linkcolor=blue,filecolor=blue,urlcolor=blue,citecolor=blue]{hyperref} 
\bibpunct{(}{)}{;}{a}{}{,}             
\makeatletter
  \newcommandtwoopt{\citeads}[3][][]{\href{http://adsabs.harvard.edu/abs/#3}%
    {\def\hyper@linkstart##1##2{}%
     \let\hyper@linkend\@empty\citealp[#1][#2]{#3}}}
  \newcommandtwoopt{\citepads}[3][][]{\href{http://adsabs.harvard.edu/abs/#3}%
    {\def\hyper@linkstart##1##2{}%
     \let\hyper@linkend\@empty\citep[#1][#2]{#3}}}
  \newcommandtwoopt{\citetads}[3][][]{\href{http://adsabs.harvard.edu/abs/#3}%
    {\def\hyper@linkstart##1##2{}%
     \let\hyper@linkend\@empty\citet[#1][#2]{#3}}}
  \newcommandtwoopt{\citeyearads}[3][][]%
    {\href{http://adsabs.harvard.edu/abs/#3}
    {\def\hyper@linkstart##1##2{}%
     \let\hyper@linkend\@empty\citeyear[#1][#2]{#3}}}
\makeatother
\usepackage{graphicx}
\usepackage{txfonts}
\usepackage{lipsum}
\usepackage{subcaption}         
\usepackage{lscape}             
\usepackage{placeins}           
\usepackage{listings}

\newcommand{\chisq} {$\chi^2_\mathrm{red}$\xspace}
\newcommand{\Msun}{\>{$\rm M_{\odot}$}\xspace}

\newcommand{\num} {$\nu_{\rm max}$\xspace}

\newcommand{\dnu} {$\Delta\nu$\xspace}

\newcommand{\Gaia} {\textit{Gaia}\xspace}

\newcommand{\GaiaDR} {\textit{Gaia}\,DR3\xspace}

\newcommand{\Kepler} {\textit{Kepler}\xspace}

\def\teff{$T_{\mathrm{eff}}$\xspace}

\def\dnu{$\Delta\nu$\xspace}
\def\dn1{$\delta\nu_{01}$\xspace}
\def\dn2{$\delta\nu_{02}$\xspace}

\usepackage{xcolor}

\begin{document}

   \title{Ensemble Individual Mode Frequency Modeling of \emph{Kepler} Giants}

    \subtitle{Extending the IACgrid towards the red-giant branch}

\titlerunning{Ensemble Individual Mode Frequency Modeling of \emph{Kepler} Giants}
\authorrunning{D.\,H. Grossmann, et al.}

\author{D.\,H. Grossmann\inst{\ref{inst:IAC},\ref{inst:ULL}} 
\and S.\,Mathur\inst{\ref{inst:IAC},\ref{inst:ULL}}
\and P.\,G. Beck\inst{\ref{inst:IAC},\ref{inst:ULL}}
\and D.\,Godoy-Rivera \inst{\ref{inst:IAC},\ref{inst:ULL}}
\and
B.\,Liagre \inst{\ref{inst:CEA_Bastien}} 
\and R.\,A.\,Garc\'ia \inst{\ref{inst:CEA}} 
\and \\
L.\,S.\,Schimak \inst{\ref{inst:Sydney}} \and
D.\,B.\,Palakkatharappil \inst{\ref{inst:CEA}} \and
M.\,H.\,Pinsonneault\inst{\ref{inst:OSU},\ref{inst:OSUCosmology}} 
\and
F.\,Pérez Hernández  \inst{\ref{inst:IAC},\ref{inst:ULL}}
} 

\institute{Instituto de Astrof\'{\i}sica de Canarias, E-38200 La Laguna, Tenerife, Spain; \label{inst:IAC}
\email{desmond.grossmann@iac.es}
\and Departamento de Astrof\'{\i}sica, Universidad de La Laguna, E-38206 La Laguna, Tenerife, Spain \label{inst:ULL}
\and Universit\'e Paris-Cit\'e, Universit\'e Paris-Saclay, CEA, CNRS, AIM, 91191, Gif-sur-Yvette, France \label{inst:CEA_Bastien}
\and Universit\'e Paris-Saclay, Universit\'e Paris Cit\'e, CEA, CNRS, AIM, 91191, Gif-sur-Yvette, France \label{inst:CEA} 
\and Sydney Institute for Astronomy (SIfA), School of Physics, University of Sydney, NSW 2006, Australia 
\label{inst:Sydney}
\and Department of Astronomy, The Ohio State University, Columbus, OH 43210, USA \label{inst:OSU}
\and Center for Cosmology and AstroParticle Physics, The Ohio State University,
191 West Woodruff Avenue, Columbus, OH 43210 \label{inst:OSUCosmology}}
   \date{Accepted on November 2, 2026}

 
  \abstract
   {Determining precise stellar parameters, in particular ages, is vital for gaining a deeper understanding of the formation and history of the Milky Way. Usually, this is achieved by fitting stellar evolution models to observational constraints such as spectroscopic and astrometric parameters. By adding the global asteroseismic properties to the classical parameters (global seismic modeling), some of the degeneracies are eliminated, and the modeling results improve. However, this approach has its limitations, with red-giant random age uncertainties usually $\sim$\,20-50\,\%, but often lacking proper quantification of the systematics.
   We aim to extend seismic modeling to include individual mode frequency modeling for red-giant branch stars, increasing the number of evolved solar-like oscillators with precise stellar parameters (ages, masses, and radii), and allowing us to quantify the systematics in global seismic modeling.
   We developed a grid of stellar evolutionary models and computed the individual oscillation frequencies, including a figure of merit to find the best model that fits the observables. We validate our methodology using a sample of red giant stars with individual frequency modeling available from the literature. Applying our methodology to this sample, we find our derived masses, radii, and ages to be in close agreement with the literature values. We then apply our method to around 100 \textit{Kepler} red-giant branch stars of approximately solar mass and metallicity, which previously have only been modeled with global seismic parameters.
   We obtain random average uncertainties of $3.4\,\%$, 1.3\,\%, and $13.2\,\%$, for stellar masses, radii and, ages, respectively. 
   We find that stellar ages from individual mode frequency modeling are significantly, 14\,\% on average, larger than those from global seismic mass-constrained modeling. The resulting modeled luminosities agree within 1.0\,\% with the luminosities acquired from \Gaia parallaxes.
    We conclude that modeling individual mode frequencies improves age precision by a factor of two for red-giant branch stars, providing stronger constraints for galactic archaeology.}


\maketitle
\nolinenumbers

\section{Introduction}
Unlocking the evolutionary history and origins of our Galaxy fundamentally depends on deriving precise stellar properties, in particular stellar ages \citep[e.g.][]{Freeman2002,Gallart2019}. However, stellar ages are only inferable from modeling age-dependent observables \citep{Agereview}. Isochrone fitting long served as a standard approach for estimating stellar ages across evolutionary stages; however, with varying precision. While for sub-giant stars close to the main-sequence turnoff, it provides reasonable precision \citep{GodoyRivera2021}, it fails to deliver that for main-sequence (MS) and red-giant branch (RGB) stars, where typical relative age uncertainties with this approach are typically 30-50\,\% \citep{Cassagrande2016}. 
Recently, space missions such as the NASA \Kepler mission \citep{Borucki2010} and the NASA Transiting Exoplanet Survey Satellite \citep[TESS;][]{Ricker2014}, have provided photometric data that allowed us to obtain precise asteroseismic parameters from observables for $\sim$$10^5$ stars on the MS and RGB presenting solar-like oscillations \citep[e.g.,][]{Yu2018, Mackereth2021,Mathur2022,Hatt2024,Lund2024,Hon2024,Nielsen2025,Hookway2025,grusnis2025}. These precise asteroseismic datasets, combined with advancements in stellar evolutionary modeling, enable a further, more precise method for age determination of MS and RGB stars: asteroseismic modeling. Ages derived from asteroseismic modeling have already proven advantageous for Galactic studies \citep[e.g.,][]{Miglio2021}.\looseness=-1

Generally, asteroseismic modeling combines spectroscopic and seismic observables to determine stellar masses, radii, and ages. Two distinct approaches are used: fitting global seismic parameters (e.g., the large frequency separation, $\Delta\nu$, and the frequency of maximum oscillation \num), or fitting individual mode frequencies.
The global seismic modeling approach has been applied successfully to several thousand MS and RGB stars, in particular thanks to projects such as the third iteration of the Apache Point Observatory \Kepler Asteroseismic Science Consortium collaboration \citep[APOKASC-3;][]{APOKASC3}. It is especially powerful when combined with constraints from binary systems \citep{Murphy2021,Grossmann2025,Schimak2026}, globular clusters \citep[e.g.][]{Moser2023,Howell2024} or open clusters \citep{Brogaard2023}. However, when applied to single field stars without further constraints available, it is usually limited to a precision of  $\approx5\,\%$ in mass and $\approx20\,\%$ in age. Furthermore, global asteroseismic age estimations are known to have systematic biases \citep[e.g.][]{tayar2025}, in particular in certain parameter ranges, such as in the low metallicity regime \citep{Epstein2014} or in binary systems due to mass loss \citep{Mazzi2025}.

Another method for determining stellar ages and masses is the modeling of the individual oscillation modes, using stellar oscillation fitting codes on top of stellar evolution models \citep[e.g.][]{Mathur2012,Metcalfe2014,LiTanda2024,Schimak2026}. As in the case of global asteroseismic modeling, it combines the seismic parameters, here the individual frequencies, with classical parameters from spectroscopy and fits them to the observables using a figure of merit. It is to date the most precise method for mass and age estimation for evolved solar-like oscillators, delivering a relative mass and age uncertainty of $\approx 2\,\%$ and $\approx 10\,\%$, respectively \citep{LiTanda2024}. However, calculating these grids, in combination with the oscillation frequencies, and the subsequent best-fit model searches tend to be very computationally expensive, especially when modeling late-type solar oscillators, such as RGB  or red-clump stars  \citep{Schimak2026}. Therefore, it has mostly been applied to MS and sub-giant branch stars so far \citep[e.g.,][]{silvaaguirre2015,SilvaAguirre2017,IACGridLucia2023,Beck2026,Garcia2026}, but more recently has been also applied to evolved stars \citep[e.g.,][]{Huber2024,Lindsay2025}.\looseness=-1

One of these modeling efforts is the IACgrid \citep{IACGridLucia2023}, a grid of stellar evolutionary models with oscillation modes to obtain precise masses, radii, and ages for target stars. However, currently, this grid stops at the base of the RGB. 
In this work, we extend the IACgrid to the RGB, develop a custom best-fit model search, and derive improved stellar parameters. We test our method on a sample of stars previously modeled using individual frequencies as constraints to validate the method. Then, we model a subsample of around 100 stars from the APOKASC-3 sample with individual frequencies, for which only global seismic modeling has been done so far. 

This paper is divided into the following sections. In Sect.\,\ref{sec:sample} we define both for our validation and main sample of red-giant-branch stars. Section\,\ref{sec:modleinggrid} explains the construction of the stellar modeling grid and computation of the individual frequencies, while Sect.\,\ref{sec:optimization} describes the search for the best-fit model and uncertainty estimation. In Sect.\,\ref{sec:validation} we apply our grid modeling to our validation sample in order to test our method. Section\,\ref{sec:results} describes the results of our method applied to the main target sample, and discusses the results compared to global seismic mass-constrained modeling and literature. The conclusions are presented in Sect.\,\ref{sec:conclusion}.
\section{Sample Selection}
\label{sec:sample}
For this work, we selected two distinct samples of RGB stars to test and apply our methodology. First, we used a set of stars that already have individual mode-frequencies modeled from the literature as our validation sample. Subsequently, we carefully selected stars with identified individual frequencies and spectroscopic parameters from the Apache Point Observatory Galactic Evolution Experiment \citep[APOGEE,][]{APOGEE}, but lacking individual mode frequency modeling, to comprise our main sample.\looseness=-1
\subsection{Validation sample}
 One of the most comprehensive studies of modeling of RGB stars, including individual mode-frequencies, can be found in \cite{Fernando2016}, which models 19 RGB stars using global parameters and individual oscillation modes \citep{Corsaro2015}.  We thus take the sample for which \cite{Fernando2016} provides $\log g$ and metallicity (16 out of 19 stars) to validate our methodology. They are shown in Fig.\,\ref{fig:sampleHRD} as red markers in the seismic Hertzsprung–Russell\,diagram (HRD).\looseness=-1
 
\subsection{Main Sample}
\label{sec:mainsample}
To select our main sample for this work, we started with the catalog of 15,808 evolved stars from APOKASC-3 \citep{APOKASC3}. We cross-matched the stars from the APOKASC-3 catalog with the sample of 6,179 RGB peak bagged stars from \cite{Kallinger2019}. To date, this catalog is the largest published collection of solar-like oscillators with peak-bagged modes, which are extracted using the Automated Bayesian peak-Bagging Algorithm (ABBA).

\cite{Beck2025} shows that the stellar-binary coevolution has a significant impact on the stellar mass and age distribution. To avoid potential binary contamination, we excluded 818 potential binary stars, flagged with the binary union flag in \cite{GodoyRivera2025}. As an additional quality cut, we only considered those stars that are part of the APOKASC-3 gold sample \citep{APOKASC3}, thereby selecting the stars with the most reliable detections from a large number of asteroseismic pipelines. This high-quality cut yielded 5,125 stars in the sample. Subsequently, we exclude known fast rotators to avoid excessive rotational splitting of the modes \citep[e.g.,][]{Beck2012,Mosser2015} and complex mixing or other divergencies caused by rotation. We did this by cross-matching the sample with the rotation catalogs from \cite{Gehan2024} and \cite{Hatt2024}. This excluded 525 stars from the sample. We furthermore excluded stars rich in $\alpha$-elements, using the $\alpha$-poor 
flag in APOKASC-3 \citep[][ 931 stars excluded]{APOKASC3}. This selection ensures a homogeneous sample consistent with the solar-scaled abundance mixture assumed in our model grid. The resulting sample has a median $[\alpha/\mathrm{Fe}]$ of 0.03\,dex, with a maximum of 0.06\,dex (typical uncertainty of 0.02\,dex), effectively indistinguishable from a purely solar-scaled composition \citep{buder2021}. Extending the analysis to $\alpha$-rich stars would require to compute a significant number of new models to span a broader range of chemical compositions along with the calculation of a significant number of individual mode frequencies. Given the computational expenses this 
would entail, such an extension is outside the scope of this work. We filtered the remaining stars, close to solar in mass and radius, in terms of their global asteroseismic properties: the frequency of maximum oscillation \num and the large frequency separation, \dnu. We only kept those targets that showed a 5\,\% or better agreement between the global asteroseismic properties as adopted in APOKASC-3 compared to those determined by \cite{Kallinger2019}.  After applying all quality cuts, this yielded a sample of 3,575 stars. 

\begin{figure}
        \centering
        \includegraphics[width=\columnwidth,height=79mm
]{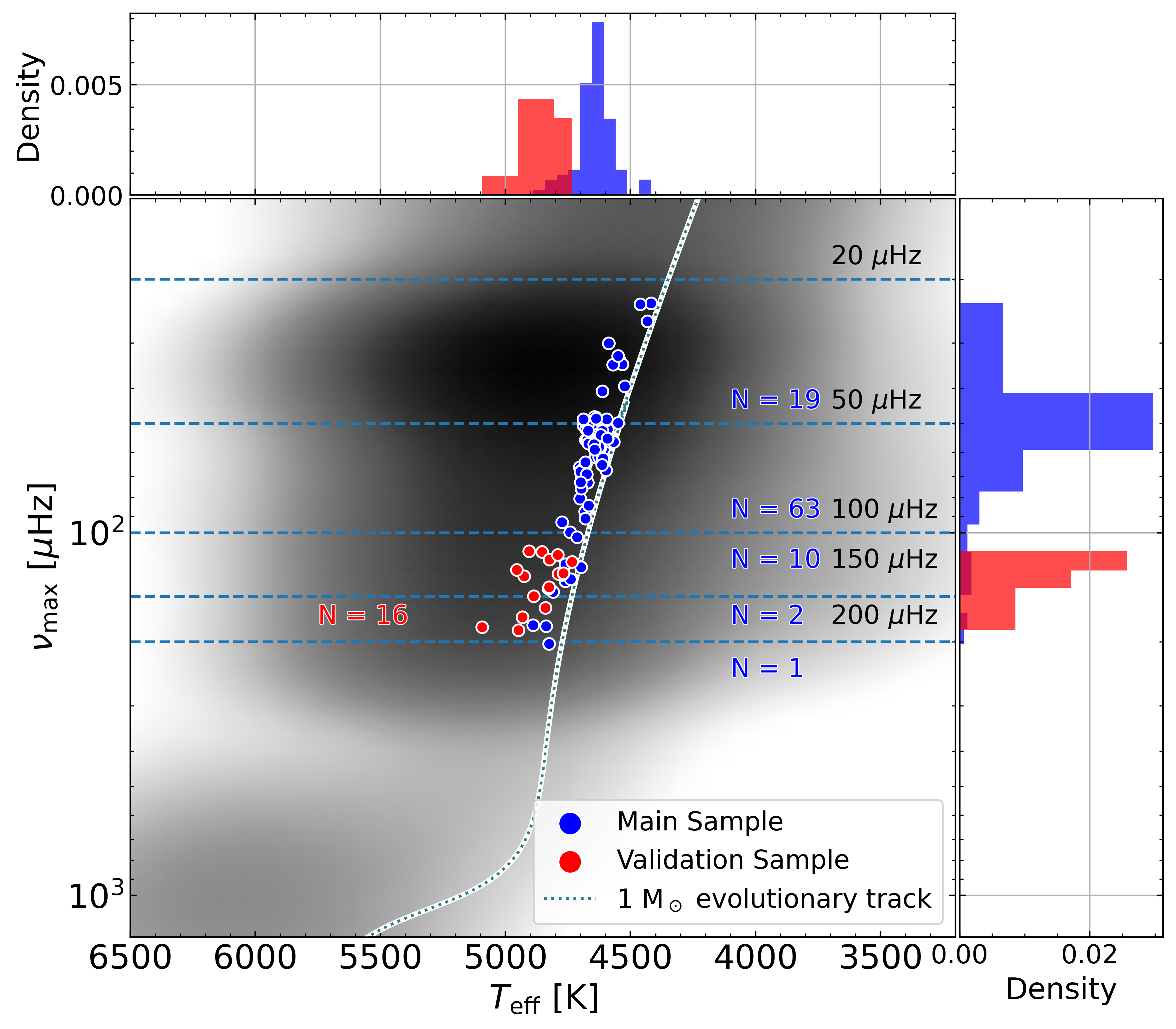}
        \caption{Seismic Hertzsprung-Russell diagram of the validation sample (red markers) and main sample (blue markers) with a representative 1.0 \Msun evolutionary track from MESA \citep[Modules for Experiments in Stellar Astrophysics;][version r23.05.1]{Paxton2011,Paxton2013,Paxton2015,Paxton2018,Paxton2019,Jermyn2023}. The background density contour plot depicts the distribution of the targets from the input catalogs used in \cite{BeckGrossmann2023, Mathur2022}. The side and top histograms show the distribution of the samples across the \teff and \num plane.\looseness=-1}
        \label{fig:sampleHRD}%
\end{figure}

As the aim of this work is to test our method on stars with similar properties as the Sun first, we only chose stars from our quality sample that have approximately solar metallicity ($[\mathrm{Fe}/\mathrm{H}] = 0.0 \pm 0.1$\,dex). Generally, the same cut could have been applied regarding stellar mass. However, in the APOKASC-3 sample, a sharp decrease in RGB targets just below 1.0 \Msun is present \citep{APOKASC3}. Therefore, we decided to center our sample at 1.1\,$\pm$\,0.1\,\Msun instead. Furthermore, as the upper part of the RGB is computationally expensive to traverse with stellar evolution and oscillation codes, we limit our sample to stars below $\log$\,(L/L$_{\odot}$)\,=\,1.8 (Sect.\,\ref{sec:modeling}). Applying these restrictions and filtering only for stars given as RGB in APOKASC-3 brought the number of targets in our main sample to 139. The complete process of sample selection is depicted in Fig.\,\ref{fig:sample}. The main sample, along with the validation sample, is depicted in the seismic HRD in Fig.\,\ref{fig:sampleHRD}.\looseness=-1

\section{The IAC red-giant Grid}
\label{sec:modleinggrid}
In this section, we describe the construction of the IAC red-giant modeling grid. It consists of a 1-D stellar evolution models, including individual oscillation frequencies calculated with a stellar oscillation code.

\subsection{Modeling Grid}
\label{sec:modeling}
For the computation of our modeling grid, we used the 1-D stellar evolution code MESA \citep[Modules for Experiments in Stellar Astrophysics;][version r23.05.1]{Paxton2011,Paxton2013,Paxton2015,Paxton2018,Paxton2019,Jermyn2023}. We calculated a grid of stellar models starting from the pre-main sequence up to a luminosity limit of $\log$ (L/L$_{\odot})$\,=\,1.8, which we used as a stopping criterion for all models. We varied the stellar mass $M$, heavy element abundance $Z$, initial helium $Y$, and mixing length parameter $\alpha$, while the parameter ranges and their rationals are described below.
 Regarding input physics and inlist settings, we generally adopted the same ones as in \cite{Grossmann2025}, which we briefly remind here. For the atmospheric boundary conditions, we chose the default Eddington-grey atmosphere $T_{\tau}$ relation \citep[default in MESA; see][]{Paxton2011} as a common choice for asteroseismic grid modeling in literature \citep[see e.g.,][]{choi2018a,Li2022surf,theodoridis2026}. Following the approach of \cite{Schimak2026} we expanded the outer limit of the MESA model to a optical depth of $\tau\,=\,10^{-3}$, to include the atmospheric data in the model and reduce surface effects when calculating stellar oscillations. We chose not to have a fixed Helium enrichment law enforced, but instead use Helium as a free parameter. Similarly, we employed classical mixing length theory \citep[MLT; e.g.,][and references therein]{JoyceTayar2023}, as defaulted in MESA \citep{Paxton2011}, and exponential core convection overshooting \citep[e.g. ][and references therein]{Constantino2015over}, with the choice of the numerical values described below. The complete MESA inlists are given in Appendix\,\ref{app:gyre}.

For the parameter of stellar mass, we had to choose a range that encompasses both our validation and main sample. The stars in the validation sample have masses ranging from 0.95 to 1.45 \Msun, while the mass range of the main sample is well within this range (1.0-1.2 \Msun). Adding a safety margin of 0.05 \Msun at both ends, we set our model mass range at 0.90 to 1.50 \Msun. As a stepsize, we chose 0.02 \Msun, as a compromise between precision and computational efficiency.

For the heavy metal abundance, to cover the parameter space given by the validation sample, we selected Z\;=\,0.004 as a lower boundary and Z\;=\,0.022 as an upper boundary, which encompasses the parameter space of the main sample as well. The stepsize in Z was set to 0.002. This choice was taken for simplicity, given the limited metallicity range (around solar) of our selected sample. We nonetheless note that for samples that span a range of metallicities, this equal spacing in Z approach would not be optimal, and instead a fixed [Fe/H] step size should be adopted.
Regarding the initial Helium abundance, we followed the reasoning in \cite{Grossmann2025} and used 0.24 and 0.30 lower and upper boundaries, respectively. For computational efficiency, we used a stepsize of 0.02. While our lower boundary of $Y = 0.24$ lies marginally below the recommended primordial helium abundance of $Y_p$\,=\,0.245\,$\pm$\,0.003 \citep{workman2022}, it remains within $1.5\sigma$ of this value and was included to ensure the grid does not impose artificial boundaries near physically plausible values.\looseness=-1 

For the mixing-length parameter, we followed the approach of \citet{Grossmann2025}, who tested different values of $\alpha_\mathrm{MLT}$ with the same input physics as used here (their Appendix\,A) before restricting the grid. Based on similar tests for our sample, we vary $\alpha_\mathrm{MLT}$ between 1.5 and 1.9 in steps of 0.2.
Regarding overshooting, we decided to use a constant value of 0.02, as in \cite{Grossmann2025}. This value is consistent with those adopted in comparable asteroseismic modeling studies of RGB stars \citep{Fernando2016, Grossmann2025}. As the overshoot length scale is defined in units of the pressure scale height $H_p$, which diverges towards the stellar center, overshooting from small main-sequence convective cores can become unphysically large. To avoid this, we use MESA's default setting, which limits the value of $H_p$ used for overshooting to the radial thickness of the convection zone divided by $\alpha_\mathrm{MLT}$ (\texttt{limit\_overshoot\_Hp\_using\_size\_of\_convection\_zone = .TRUE.}).
The parameters of the modeling grid are summarized in Table \ref{tab:parameters}.
\begin{table}[t!]
\caption{Parameter range for model grid used in this work.}
\tabcolsep=6.5pt 
\centering
\begin{tabular}{lccc}
\hline
\hline
Parameter &  Unit & Value (range)  & Step size    \\ \hline
Mass (M)                & [\Msun] & [0.90; 1.50]   & 0.02 \\

Initial Metallicity ($Z$) & $-$ & [0.004; 0.022]    & 0.002   \\
Initial helium ($Y$) & $-$ & [0.24; 0.30]& 0.02 \\
Mixing length ($\alpha_\mathrm{MLT}$) & $-$ & [1.5; 1.9] & 0.2 \\
Overshooting ($f_{\mathrm{ov}}$) & $-$  & 0.02 & $-$ \\ 

\hline            
\end{tabular}

\label{tab:parameters}

\tablefoot{The first two columns define the parameter and its unit used for the grid modeling. The third and fourth columns define the grid's value range and step size. The last row shows the overshooting Parameter $f_{\mathrm{ov}}$, which was kept constant for all models.}
\end{table} 

\subsection{Computation of individual frequencies}
For the MESA stellar evolutionary models described in the previous section, we calculated the radial and quadruple oscillation modes using the stellar oscillation code GYRE \citep[][version 7.1]{Townsend2013}. For computational efficiency, we choose to only compute oscillation modes for every fifth stellar model, provided by the MESA custom time step selection \citep{Paxton2011}, and limit the region for mode estimation to \num $\pm$4 \dnu of the respective stellar model. This was further motivated by the fact that most of the stars had around 7 consecutive orders of modes observed. For the radial modes, we chose a resolution of the frequency scanning grid (\texttt{n\_freq}) of 100, which would cover all potential radial modes in the given interval. Regarding the quadruple modes, the number of observed modes in any given range is equal to or very similar to the number of radial modes. Although the observed quadrupole modes are nearly pure pressure modes, the coupling between the stellar core and envelope gives rise to a much larger set of mixed modes. All of these mixed modes are computed by GYRE. However, due to their high inertia, the majority of mixed quadrupole modes are not observed in the power spectrum \citep{Dupret2009}. For this reason, we increased the frequency scanning resolution for the quadruple modes significantly to \texttt{n\_freq\,=\,400}, to ensure not to miss any relevant modes. Subsequently, we selected only those modes corresponding to the local inertia minima, using the \texttt{scipy} \citep{2020SciPy-NMeth} \verb|find_peaks| method for our further analysis.\looseness=-1

We initially considered including dipole modes in the modeling effort as well. However, visual inspection of the observed $\ell$=1 modes revealed a significant amount of mixing and rotational splitting, even after excluding known fast rotators (Sect.\,\ref{sec:mainsample}). This would pose significant challenges in modeling and exceeds the scope of this work. Therefore, we decided not to include them as constraints to find the best-fit models. The full inlist file used for the mode calculation is given in Appendix \ref{app:gyre}.

The modes for every 5th model timestep were calculated as described and subsequently interpolated over time in order to have theoretical oscillation modes for every timestep in our stellar modeling. We tested several interpolation methods as implemented in the \texttt{scipy} package, comparing their results with the true oscillation values from GYRE. From our test, we found that the Piecewise Cubic Hermite Interpolation (PCHIP) method was the most suited to our problem as it gave us the least deviation from the true frequency values with little to no oscillatory behaviour.\looseness=-1

\subsection{Surface Correction} \label{sec:surfcorr}
In stars with solar-like oscillations, a known discrepancy exists between the modeled and observed frequencies. This is known as surface effects, since it is believed to emerge from inaccurate modeling of the surface layers of stars in 1D-stellar modeling.\looseness=-1

To correct for these surface effects that current stellar oscillation codes do not account for, different corrections have been proposed in the literature \citep[e.g.,][]{Kjeldsen2008,BallGizon2014,Roxburgh2016}. We used the two-term formalism from \cite{BallGizon2014}, as it is the most well-tested one for evolved stars \citep[][and references therein]{Schimak2026}. For a given oscillation mode at a frequency $\nu$ with an Inertia $\mathcal{I}$ the correction is given as,
\begin{equation}
\delta\nu=\left(a_{-1}(\nu/\nu_{ac})^{-1} +a_3(\nu/\nu_{ac})^3\right)/\mathcal{I}\text{,}
	\end{equation}
where $ a_{-1} $ and $ a_{3} $ are coefficients to be fitted for a given target, and $\nu_{ac}$, the acoustic cutoﬀ frequency. The latter is calculated using the surface gravity $g$ and the effective temperature of the model $T_{\text{eff}}$ in the scaling relation \citep{BallGizon2014},
\begin{equation}
\nu_\text{ac}/\nu_{\text{ac},\odot}=\frac{\text{g}}{\text{g}_\odot}\left( \frac{{\text{T}_{\text{eff}}}}{\text{T}_{\text{eff},\odot}}\right)^{-1/2},
	\end{equation}
where we take $\nu_{\text{ac},\odot}=5000\,\mu Hz$ \citep{BallGizon2014}.  The $ a_{-1} $ and $ a_{3} $ parameters are determined for each timestep of each model separately by using the procedure described in Sect. 2.1 of \cite{BallGizon2014}. As this correction relies on a simple scaling law, we may apply an overcorrection and hence fit a non-physical stellar (oscillation) model to the observed mode spectrum. Therefore, we compared the estimated surface correction at \num with literature values for RGB stars. Most relative surface corrections are expected to be in the range of 0.5\,\% and 1\,\% of \num and have a negative sign \citep[][and references therein]{Li2022surf,Wang2023}. For all stars in our sample the correction falls in the margin of 2\,\% of \num used in \cite{Li2022surf}, with most of them below 1\,\%. 

We also tested for a potential correction of the frequencies for Doppler velocity shifts \citep{davies2014}. However, our tests, using APOGEE mean radial velocities, show that for our target stars these Doppler corrections amount to about 0.05\,\% to 0.1\,\% of the frequencies at \num, hence at least one order of magnitude smaller than the surface corrections and usually also smaller than the uncertainties reported on the mode frequencies. As in our case, the surface correction would mitigate this Doppler correction, so we decided not to explicitly correct the observed frequencies for it.

\section{Best-fit model search}
\label{sec:optimization}

In this section, we define the procedure for finding the best-fit model. Furthermore, we describe the input parameters used in the best-fit model search and the uncertainty estimation.

\subsection{Definition of figure of merit function}
For each timestep of the models (Sect. \ref{sec:modeling}), we used a figure of merit function to find the model that best matches the observables (Sect. \ref{sec:sample}) of the given target.
As described in \cite{Grossmann2025} we used a reduced \chisq-minimization approach given as,
\begin{equation}
    \chi^2_\mathrm{red} = \frac{1}{N-m}\sum_{k=1}^{N} \left( \frac{(y_{k,\text{obs}}-y_{k,\text{mod}})}{\sigma_k} \right)^2\,,
    \label{eq:chitot}
\end{equation}
where $y_{k,\text{obs}}$ represents the observed quantity and $ y_{k,\text{mod}}$ the respective quantity from modeling, with $\sigma_k$ being the observational error. The sum is done over the number of observables $N$. The result of this sum is normalized by dividing by the difference between $N$ and $m$, the number of fitted parameters. The $m = 7$ fitted parameters are the initial mass $M$, metallicity $Z$, initial helium abundance $Y$, mixing length $\alpha_\mathrm{MLT}$, the model age, and the two surface-correction coefficients. Since $N$ and $m$ are identical for all models of a given target, the normalization by $N-m$ does not affect the identification of the best-fit model.

\subsection{Observables included in the figure of merit}
With the general figure of merit function defined in the previous section, we now summarize the observables included in our best-fit model search.
Those parameters fall into two categories: the spectroscopic observables and the seismic observables. We decided to include the effective temperature, $T_{\mathrm{eff}}$, and the heavy element abundance, $Z$, as spectroscopic observables. For the conversion of the heavy element abundance from [Fe/H] to $Z$, we used the solar reference as given by \cite{GrevesseSauval1998}. We retrieve the observational spectroscopic parameters directly from \cite{Fernando2016} for our validation sample and from the APOKASC-3 \citep{APOKASC3} for our main sample. This gives the spectroscopic-$\chi^{2}$ defined as,\looseness=-1
\begin{equation}
\label{eq:chi spec}
    \chi^{2}_{\mathrm{spec}} = \chi_{T_{\text{eff}}}^{2} + \chi_{Z}^{2}.
\end{equation}

Regarding the seismic observables, we used the seismic surface gravity $\log\,g_{\mathrm{seis.}}$
as defined in the APOKASC-3 catalog \citep{APOKASC3} and the radial and quadruple modes as constraints. The oscillation modes together with the mode identification were extracted from \cite{Corsaro2015} for the validation sample and from \cite{Kallinger2019} for the main sample. To reduce the influence of potential outlier modes at the edges of the power excess, we choose to include only modes within $\pm 40\,\%$ of \num into the best-fit model search. Therefore, our seismic figure of merit function consists of three components, as given by,\looseness=-1 
\begin{equation}
\label{eq:chiind}
    \chi^{2}_{\mathrm{seis}} = \chi_{\log \mathrm{g}}^{2} +\chi_{l=0}^{2} + \chi_{l=2}^{2},
\end{equation}
yielding the full figure of merit function to be,
\begin{equation}
\label{eq:chifull}
    \chi^{2}_{\mathrm{tot}} =  \chi^{2}_{\mathrm{spec}} + \chi^{2}_{\mathrm{seis}}.
\end{equation}
This $\chi^{2}_{\mathrm{tot}}$ is reduced by the first term in Eq.\,\ref{eq:chitot}, and used as the final figure of merit
\subsection{Best-fit model search using global seismic constrains}
\label{sec:globparmdef}
To compare our individual frequency modeling results, we perform a best-fit model search over our existing model grid using the stellar mass derived from global seismic scaling relations as reported in \cite{APOKASC3} as the seismic constraint, together with spectroscopic constraints on surface gravity and metallicity, as described in Sect.\,3.3.2 of \cite{APOKASC3}. This search therefore uses the APOKASC-3 seismic mass directly as input, rather than \dnu\ or \num, and is intentionally designed to mimic the APOKASC-3 methodology as closely as possible, while avoiding the need to compute \dnu\ from the grid or to adopt a solar \dnu\ calibration. Stellar ages and radii are then derived from the resulting best-fit model. This yields a global $\chi^{2}_{\mathrm{glob}}$ of\looseness=-1
\begin{equation}
\label{eq:chiglob}
    \chi^{2}_{\mathrm{glob}} =  \chi^{2}_{\mathrm{M}_{\mathrm{seis.}}} + \chi^{2}_{\mathrm{log\,g}} + \chi_{Z}^{2}.
\end{equation}

\subsection{Uncertainty estimation}

To compute the random uncertainties associated with the stellar parameters retrieved for the best-fit model, we apply the statistical Monte-Carlo method \citep{MonteCarlo1949}. For each target, we created a multi-dimensional parameter space of normal distributions based on the observational input parameters and their 1$\sigma$ uncertainties. From this parameter space, we randomly draw $N_{\mathrm{MC}}$ points and use them as new input vectors for our figure of merit. Since uncertainties in the input parameters propagate to the age in a highly non-linear manner \citep[e.g.][]{hon2020}, the resulting age distribution does not in general approach a normal distribution even at large $N_{\mathrm{MC}}$. Instead, $\log(\mathrm{age})$ more closely follows a normal distribution, as the dominant nonlinearities propagate to age as a power law. The Monte Carlo approach is particularly suited to this problem as it simultaneously considers errors in all dimensions of the observational parameter space, naturally accounting for degeneracies and nonlinearities in the mapping from observables to stellar age (or other stellar parameters; \citealt{jorgensen2005,pont2004}). In general, similar approaches are also  commonly used in determining uncertainties in asteroseismic (grid) modeling \citep[e.g.][]{serenelli2017a,borre2022, Verma2022Helium}.\looseness=-1

With few exceptions, the median of the output distribution was very close (< 7\,\%) to the best-fit values, which further indicates that the output distribution is reliable. As commonly used in the literature \citep[e.g.][]{Godoy25}, we adopted the 16\textsuperscript{th}- and 84\textsuperscript{th}-percentile of the output distributions as the lower and upper uncertainties of age, mass, and radius in our modeling, respectively. For consistency, we used the median (50\textsuperscript{th}-percentile) instead of the mean as central values in our work.\looseness=-1

Regarding the number of iterations, we tested for a subsample of 10 targets four different numbers of iterations,  $N_{\mathrm{MC}}$\,=\,200,\,500,\,1000 and \,1500, in order to find a compromise between precision and computation time.  Our testing concluded that from $N_{\mathrm{MC}}= 500$ onwards the output median, 16\textsuperscript{th}-, and 84\textsuperscript{th}-percentiles converge with changes of around 2\,\%. Hence, we decided to use 500 iterations for the Monte-Carlo uncertainty estimation for the full sample. On average, the Monte-Carlo run using $N_{\mathrm{MC}}= 500$ for one star comprised around 10 hours of computation time. We note that mass is an inherently discrete parameter in our modeling grid, as each Monte Carlo draw maps to one of the fixed grid mass values. Since percentiles are well-defined for discrete empirical distributions, and the mass step size is smaller than the typical mass uncertainty, the $N_\mathrm{MC} = 500$ draws span a sufficient number of discrete values to yield stable and meaningful credibility intervals.

\subsection{Sample comparisons}
\label{sec:significance}
To compare stellar parameters obtained by two different modeling efforts at the sample level, we use two  quantities, namely the median deviation, and the significance of the deviation given the sample ($\Sigma$). We characterize the latter by the offset relative to the uncertainty on the sample mean, defined below.\looseness=-1

We consider the $N$ stars that are common to the two methods. For star $i$, $x_i$ and $y_i$ denote the estimates of a given parameter (e.g., age). The fractional deviation, normalized to the mean of the two estimates, is
\begin{equation}
\delta_i = \frac{x_i - y_i}{\tfrac{1}{2}(x_i + y_i)}.
\label{eq:deviation}
\end{equation}
As the uncertainties on $x_i$ and $y_i$ are asymmetric, we use the uncertainties corresponding to the direction of their difference when propagating the error on $\delta_i$. Specifically, if $x_i$>$y_i$, we combine the lower uncertainty on $x_i$ with the upper uncertainty on $y_i$; if $x_i$<$y_i$, we instead combine the upper uncertainty on $x_i$ with the lower uncertainty on $y_i$. We denote the resulting uncertainty on the fractional deviation as $\sigma_{i, f}$. Whether two modeling methods differ systematically is a property of the sample mean, whose uncertainty decreases with sample size and is thus smaller than that of an individual star. Propagating the individual uncertainties $\sigma_{i,f}$ to the mean deviation and expressing the latter in these units gives
\begin{equation}
\Sigma = \frac{\langle \delta \rangle}{\sigma_{\langle \delta \rangle}},
\qquad
\langle \delta \rangle = \frac{1}{N}\sum_i \delta_i,
\;\;
\sigma_{\langle \delta \rangle} = \frac{1}{N}\sqrt{\sum_i \sigma_{i,f}^{2}}.
\label{eq:significance}
\end{equation}
where $\langle \delta \rangle$ is the sample mean of the deviations and $\sigma_{\langle \delta \rangle}$ its uncertainty, propagated from the individual measurement errors. $\Sigma$ is the number of standard errors by which the sample offset differs from zero, and we take $|\Sigma| \gtrsim 3$ as a statistically significant offset. As it scales as $\sqrt{N}$, even a small offset can become significant in a given sample.

\section{Validating the method}
\label{sec:validation}
\begin{figure}
        \centering
     \includegraphics[width=\columnwidth,height=80mm
]{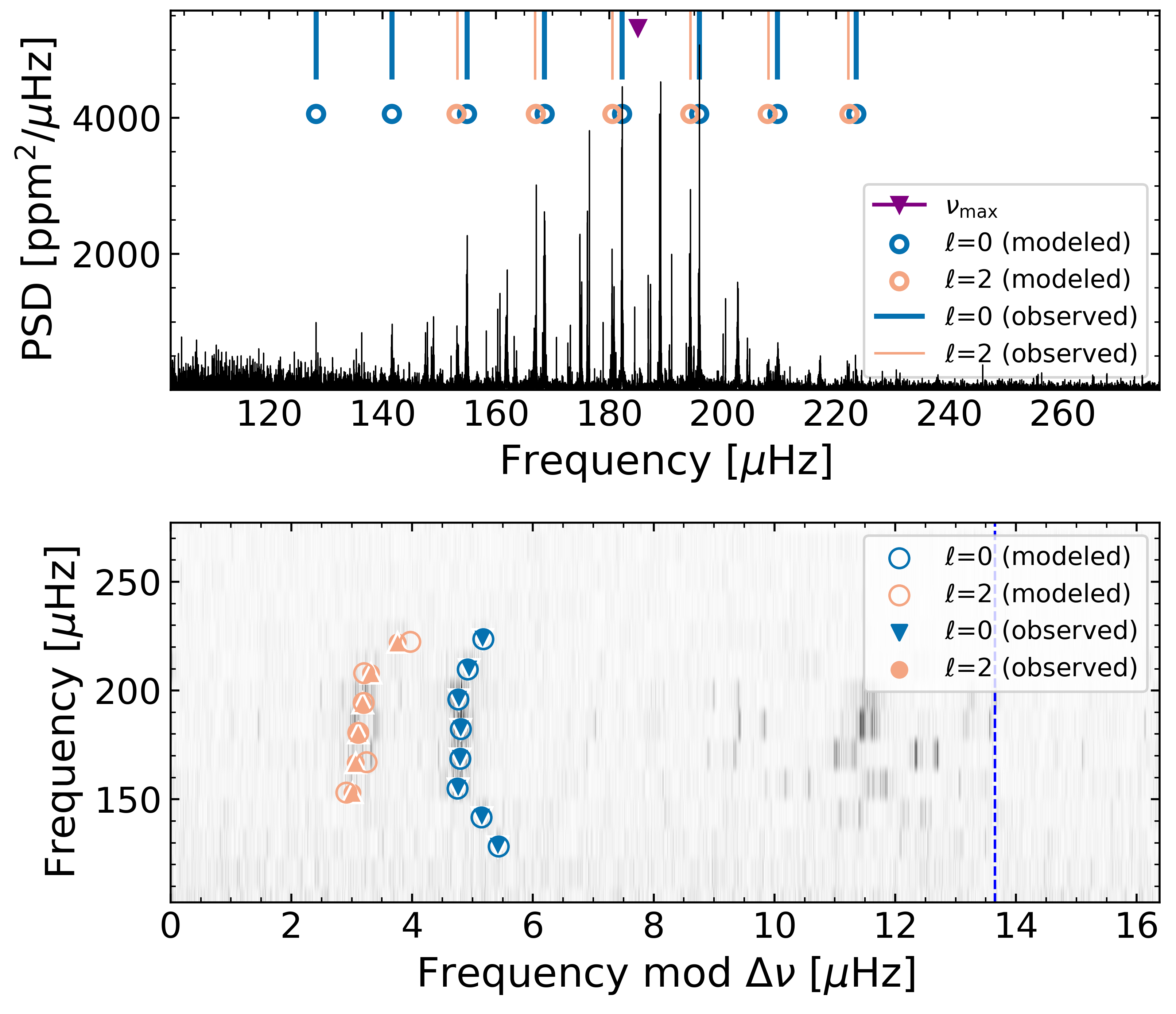}
        \caption{Top panel shows power spectral density (PSD) centered on \num for KIC\,8366239 with the modes from \cite{Corsaro2015} marked as vertical colored lines according to the legend. The best fit modes from modeling, after correction of surface effects, are given as open dots. Bottom panel shows an échelle diagram for the same target, with the observed modes marked as closed symbols and the modes from the best fit modeling as open symbols, with \dnu marked as a blue vertical line.}
        \label{fig:echellefernando}%
        \centering

\end{figure}

\begin{figure*}[t]  
    \centering
    \includegraphics[width=0.33\textwidth,height=65mm]{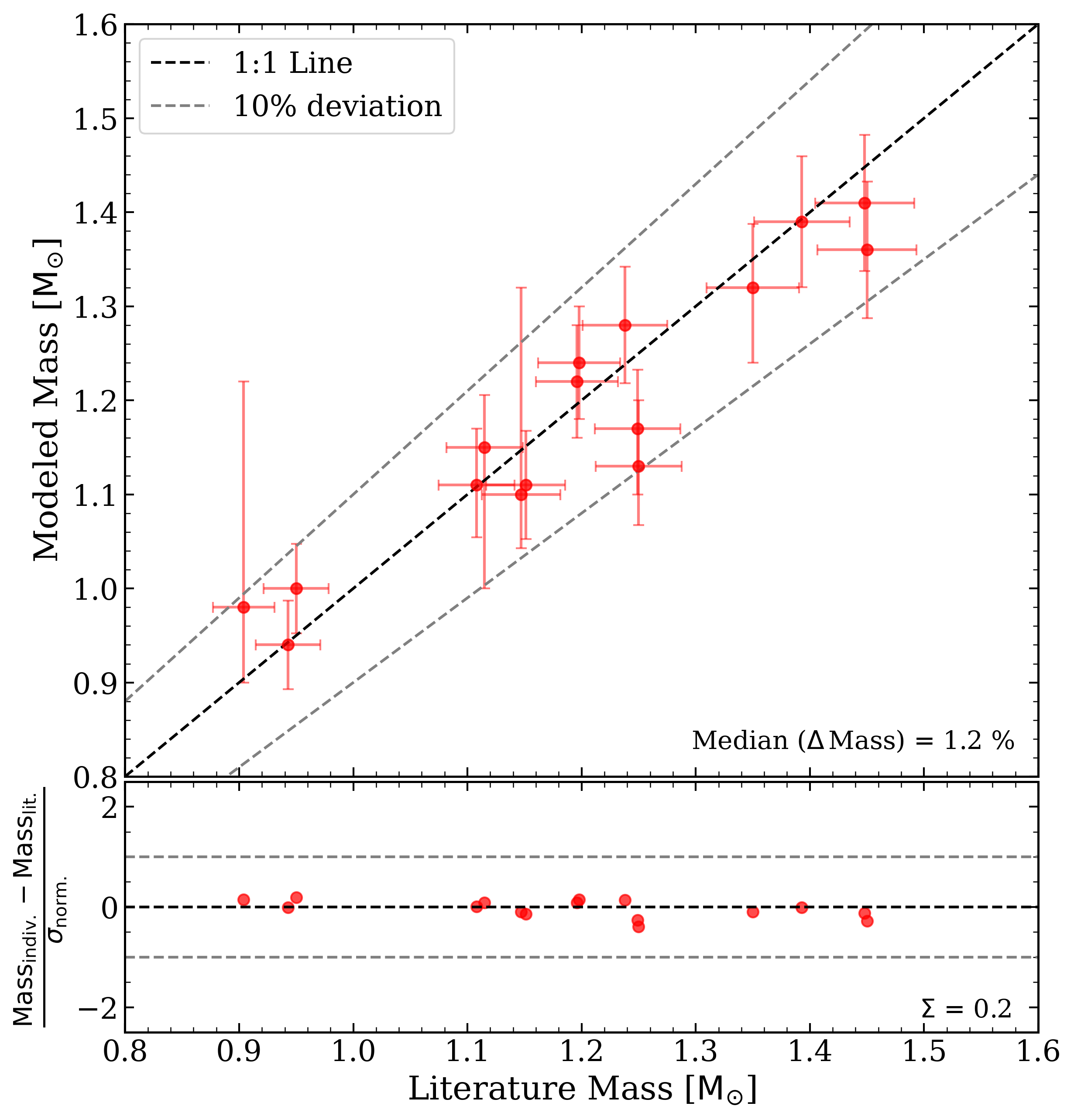}%
    \includegraphics[width=0.33\textwidth,height=65mm]{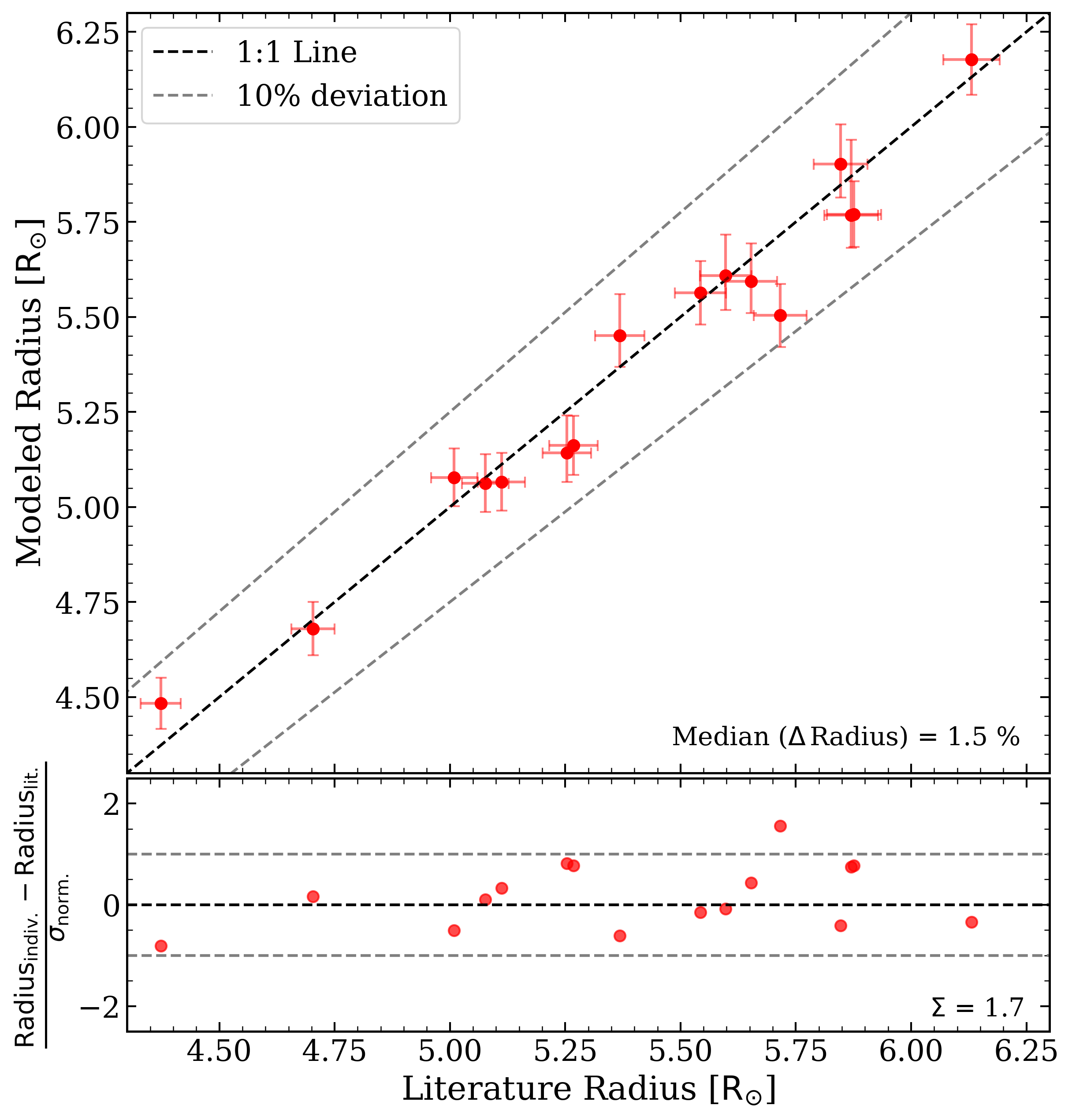}%
    \includegraphics[width=0.33\textwidth,height=65mm]{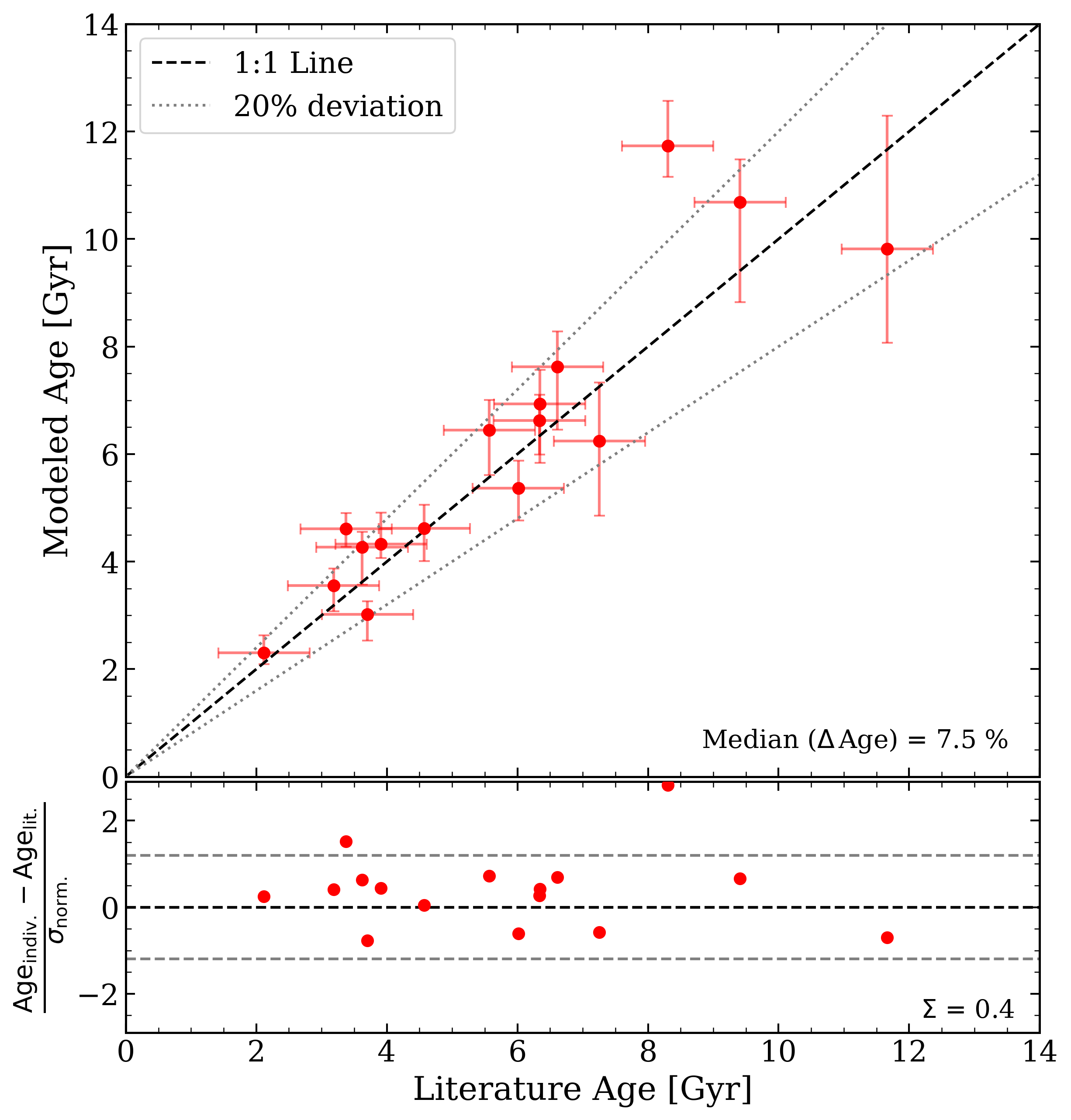}
    \caption{Comparison of mass (left top panel), radius (middle top panel), and age (right top panel) for the validation sample with the respective parameters from previous individual mode frequency modeling \citep{Fernando2016}. The respective bottom panels show the residuals, in terms of $\sigma$, between the modeled parameters from this work and the literature as a function of the literature parameter. The dashed line and dotted grey line represent a 1$\sigma$ deviation.\looseness=-1}
    \label{fig:fernando_comparison}
\end{figure*}

We first applied our method to the 16 stars of our validation sample from \cite{Fernando2016}.  An example échelle diagram, including the modeled and observed modes, is shown in Fig.\,\ref{fig:echellefernando}. Fig.\,\ref{fig:fernando_comparison} shows the comparisons of the masses, radii, and ages from our modeling with the values given in \cite{Fernando2016}. Overall, we report a median uncertainty of 3.6\,\% (0.04 \Msun) for mass, 1.7\,\% in radii, and 13.0\,\% (0.64 Gyr) for age, summarized in Table\,\ref{tab:uncertainties}. All the targets fall well within 10\,\% of the literature mass, with the median deviation from the literature values of 1.2\,\%. The mass deviation between the two methods has a sample level significance of only $\Sigma$\,= 0.2. In  case of the stellar radii, the median deviation compared to literature values is 1.5\,\%, with a $\Sigma$\,= 1.7 significance. Regarding the age, the average deviation between our modeling results and the literature values from individual mode-frequency modeling is 7.5\,\%, with a significance of $\Sigma$\,= 0.4. The comparison of the output and input effective temperatures shows an average deviation of  0.60\,\% ($\Sigma$\,= 0.4). For the surface gravity  $\log\,g$, we find a 0.007\,dex deviation on average, about half of the input uncertainty. Conclusively, we do not observe any significant, systematic biases in mass, age, or temperature compared to the values from \cite{Fernando2016}.\looseness=-1

\begin{table}
\centering
\caption{Median random uncertainties for stellar parameters from individual mode frequency modeling.}
\label{tab:uncertainties}
\begin{tabular}{lccc}
\hline\hline
Sample & $\sigma$ Mass [\%] & $\sigma$ Radius [\%] & $\sigma$ Age [\%] \\
\hline
Validation sample & 3.6 & 1.7 & 13.0 \\
Main sample       &   3.4  &   1.3  &  13.2  \\
\hline
\end{tabular}
\vspace{-5mm}
\end{table}
We note that one target, KIC\,10200377 has excellent agreement in mass (modeled mass equals the literature mass), however, the age is around 50\,\% higher than the one determined by \cite{Fernando2016}. The error analysis revealed that this is likely due to the fact that the observed metallicity \citep[Z\,=\,0.005,][]{Fernando2016} is very close to the lower edge of the grid, preventing us from finding the correct best-fit model, thus the metallicity of the target might be below our grids lower boundary of Z\,=\,0.004. It should be furthermore noted that the age uncertainties in \cite{Fernando2016} are given as one value (0.56\,Gyr) in common for all targets, hence for targets with older ages, this might lead to an underestimation of the underlying uncertainty. 

Additionally, although the modeling grid in \cite{Fernando2016} is similar to the one in this work, the figure of merit differs slightly, as it uses a different surface effect correction description and also includes the asymptotic period spacing $\Delta\Pi$, which can lead to differences in the results.
Regarding the oscillation modes, we note good agreement between the observed modes and the modes from the best-fitting modes, with the \chisq from the frequencies always below 2.
A summary of the stellar parameters obtained for the validation sample is shown in Table\,\ref{tab:resultsvalidation}.
\begin{table}[h]
\centering
\caption{Stellar parameters and random uncertainties from individual mode frequency modeling for the validation sample.}
\small
\setlength{\tabcolsep}{2pt}
\renewcommand{\arraystretch}{1.3}
\begin{tabular}{lccccc}
\hline\hline
KIC & Age$_{\rm mod.}$ & Mass$_{\rm mod.}$ & $T_{\rm eff, mod.}$ & $\log g_{\rm mod.}$ & Radius$_{\rm mod.}$ \\
 & [Gyr] & [M$_\odot$] & [K] & [dex] & [R$_\odot$] \\
\hline
3744043 & $6.45^{+0.40}_{-0.68}$ & $1.10^{+0.22}_{-0.07}$ & $4965^{+30}_{-59}$ & $3.07^{+0.02}_{-0.11}$ & $5.77^{+0.20}_{-0.07}$ \\
6117517 & $5.37^{+0.51}_{-0.42}$ & $1.24^{+0.06}_{-0.07}$ & $4773^{+30}_{-160}$ & $2.99^{+0.02}_{-0.02}$ & $5.90^{+0.10}_{-0.06}$ \\
6144777 & $6.24^{+1.02}_{-1.39}$ & $1.15^{+0.04}_{-0.15}$ & $4685^{+41}_{-30}$ & $3.06^{+0.02}_{-0.04}$ & $5.45^{+0.11}_{-0.07}$ \\
7619745 & $3.55^{+0.31}_{-0.47}$ & $1.36^{+0.06}_{-0.07}$ & $4920^{+57}_{-91}$ & $3.15^{+0.02}_{-0.02}$ & $5.16^{+0.05}_{-0.06}$ \\
8366239 & $3.02^{+0.25}_{-0.49}$ & $1.41^{+0.06}_{-0.07}$ & $4950^{+40}_{-85}$ & $3.18^{+0.06}_{-0.05}$ & $5.07^{+0.03}_{-0.03}$ \\
8475025 & $4.27^{+0.29}_{-0.70}$ & $1.28^{+0.06}_{-0.06}$ & $4859^{+59}_{-96}$ & $2.96^{+0.02}_{-0.02}$ & $6.18^{+0.07}_{-0.06}$ \\
8718745 & $10.69^{+0.79}_{-1.86}$ & $1.00^{+0.05}_{-0.04}$ & $4744^{+117}_{-46}$ & $3.02^{+0.02}_{-0.02}$ & $5.08^{+0.04}_{-0.04}$ \\
9145955 & $4.33^{+1.13}_{-0.26}$ & $1.22^{+0.02}_{-0.05}$ & $4923^{+107}_{-122}$ & $3.03^{+0.02}_{-0.02}$ & $5.56^{+0.04}_{-0.07}$ \\
9267654 & $7.62^{+0.60}_{-1.17}$ & $1.11^{+0.06}_{-0.05}$ & $4810^{+30}_{-48}$ & $2.98^{+0.02}_{-0.02}$ & $5.61^{+0.11}_{-0.09}$ \\
9475697 & $6.93^{+0.50}_{-1.09}$ & $1.11^{+0.04}_{-0.07}$ & $4759^{+30}_{-60}$ & $2.96^{+0.02}_{-0.02}$ & $5.77^{+0.05}_{-0.04}$ \\
9882316 & $2.31^{+0.54}_{-0.05}$ & $1.39^{+0.08}_{-0.09}$ & $5054^{+108}_{-99}$ & $3.17^{+0.02}_{-0.02}$ & $5.06^{+0.05}_{-0.04}$ \\
10123207 & $9.82^{+2.48}_{-1.63}$ & $0.98^{+0.24}_{-0.08}$ & $4853^{+30}_{-30}$ & $3.14^{+0.02}_{-0.02}$ & $4.48^{+0.06}_{-0.07}$ \\
10200377 & $11.74^{+0.70}_{-0.58}$ & $0.94^{+0.04}_{-0.05}$ & $4836^{+122}_{-115}$ & $3.07^{+0.02}_{-0.02}$ & $4.68^{+0.05}_{-0.06}$ \\
10257278 & $6.62^{+0.48}_{-0.63}$ & $1.17^{+0.06}_{-0.07}$ & $4812^{+91}_{-49}$ & $3.08^{+0.02}_{-0.02}$ & $5.14^{+0.10}_{-0.06}$ \\
11353313 & $4.61^{+0.29}_{-0.26}$ & $1.13^{+0.07}_{-0.05}$ & $4924^{+127}_{-36}$ & $3.01^{+0.02}_{-0.02}$ & $5.50^{+0.08}_{-0.06}$ \\
11968334 & $4.62^{+0.44}_{-0.61}$ & $1.32^{+0.07}_{-0.08}$ & $4816^{+50}_{-134}$ & $3.06^{+0.02}_{-0.02}$ & $5.59^{+0.10}_{-0.07}$ \\
\hline
\end{tabular}

\label{tab:resultsvalidation}
\end{table}
\section{Results}
\label{sec:results}
After validation of our method using a test sample, we apply it to our main sample of red-giants and discuss the results compared to the global seismic mass-constrained modeling (Sect.\,\ref{sec:globresults} and \ref{sec:indiv}). We also discuss how the results are impacted by adding the bolometric luminosity from \Gaia \citep{GaiaMain} as an independent constraint into the figure of merit (Sect.\,\ref{sec:lum}). Then, we compare our results to those derived in the literature (Sect.\,\ref{sec:lit}). 
\subsection{Global seismic mass-constrained modeling}
\label{sec:globresults}
First, we obtained the stellar parameters by selecting the best-fit model from our 
grid using the APOKASC-3 seismic mass together with spectroscopic constraints, 
as described in Sect.\,\ref{sec:globparmdef}.
The metallicities and effective temperatures of the best-fit models were all within  1$\sigma$ of the observed values. 
\begin{figure}[ht!]
        \centering
        \includegraphics[width=\columnwidth
]{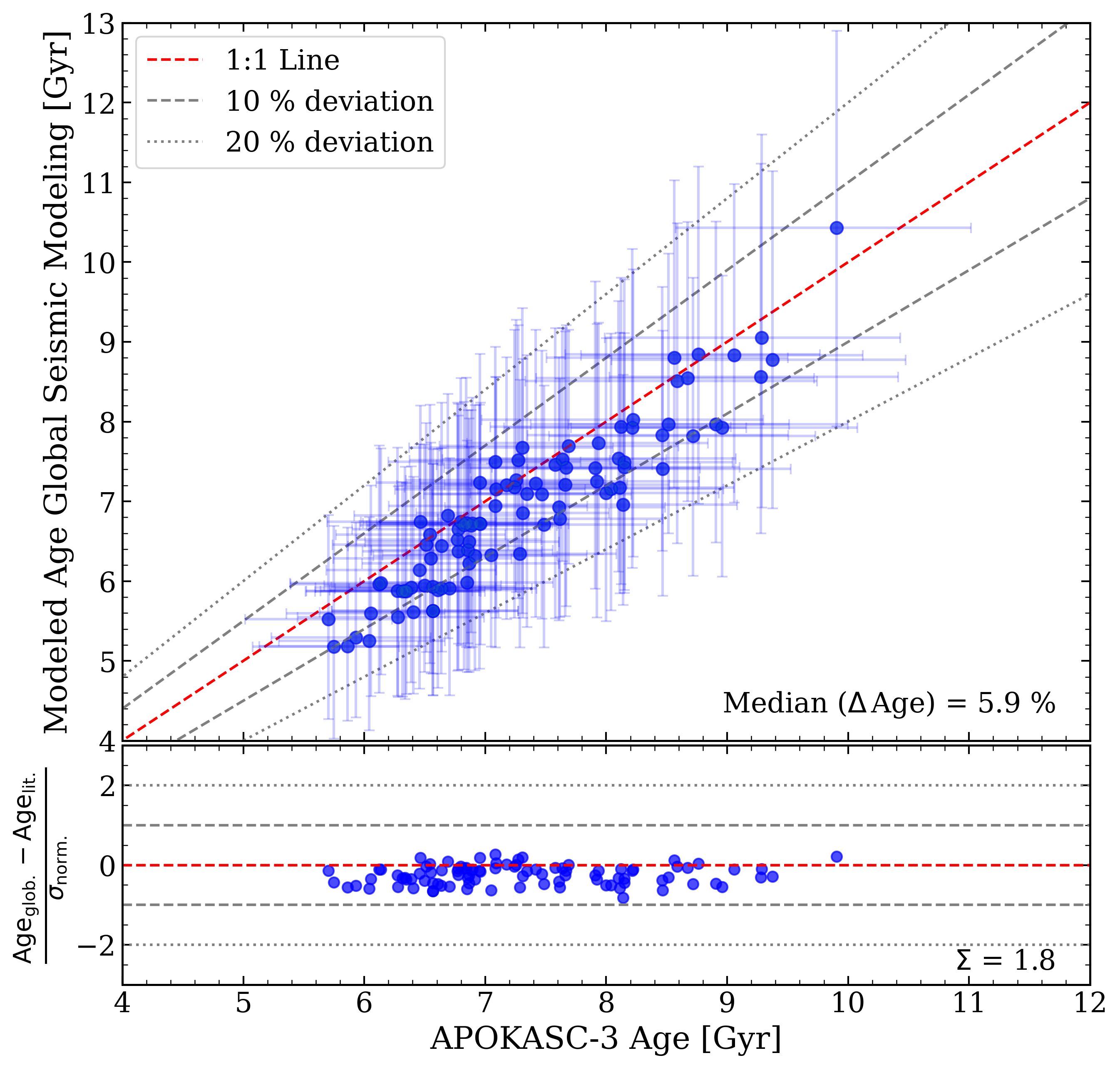}
        \caption{Top panel shows age comparison of ages from APOKASC-3 and the global seismic mass-constrained modeling from this work. Bottom panel shows residuals, in terms of $\sigma$, between the two age estimations as a function of age from APOKASC-3. The dashed line and dotted grey lines represent 1$\sigma$ and 2$\sigma$ deviation, respectively.\looseness=-1}
        \label{fig:apoglob}%
\end{figure}

Figure\,\ref{fig:apoglob} illustrates the comparison between the ages from our global modeling with the ones from the APOKASC-3 catalog. We notice a very good agreement with a median difference of 0.4\,Gyr (5.9\,\%). Part of this difference may stem from the different metallicity scales adopted in the two approaches due to the $\alpha$-element abundance correction, as APOKASC-3 applied the \citet{salaris1993} approximation to correct $[\mathrm{Fe/H}]$ for alpha-element enhancement. Nevertheless, the observed offset is well within the systematic age differences at the 5-10\,\% level seen between different calibrations within APOKASC-3 \citep{APOKASC3} itself and between APOKASC-3 and \citet{Yu2018}. Theses systematic differences are an intrinsic feature of grid-based RGB age determination \citep{silvaaguirre2020}, as figures of merits, input physics and numerical recipes can differ between methods. Furthermore,  the deviation in age between the literature values and our global seismic mass-constrained modeling has a  significance of $\Sigma$\,=\,1.8, indicating that this difference is not significant on a sample level.

Regarding stellar radii, we derive an average deviation of 0.37\,\% between the APOKASC-3 radii and our global modeling radii, corresponding to a significance of $\Sigma$\,= 1.2. For stellar mass, we obtained an average deviation of 0.01\,\Msun (0.9\,\%) from the masses in APOKASC-3 and attribute formal uncertainties of 3.7\,\% and 4.3\,\%, for lower and upper errors, respectively. However, we note that mass was an explicit input quantity in this estimation, therefore those values and uncertainties should not be regarded as independent. 

We obtain a median random uncertainty in age of 21.5\,\% and 26.0\,\%, for the lower and upper error, respectively. One relevant contribution to these random uncertainties likely stems from degeneracies involving the initial helium mass fraction $Y$ and the mixing length parameter $\alpha_\mathrm{MLT}$. The output distributions of both parameters are approximately flat across the explored grid ranges, indicating that the individual frequency data do not strongly constrain either quantity independently. Both distributions show slight enhancements at the grid boundaries (around 10 \% higher), consistent with the grid-boundary bias identified by \cite{Valle2024}, who demonstrate that poorly constrained parameters tend to be biased towards the edges of the explored parameter space regardless of the fitting method. Since age is sensitive to both $Y$ and $\alpha_\mathrm{MLT}$ at fixed mass and metallicity, this lack of constraint directly propagates into the age uncertainty.

\subsection{Individual mode frequency modeling}
\label{sec:indiv}
\begin{figure}
        \centering
     \includegraphics[width=\columnwidth,height=80mm
]{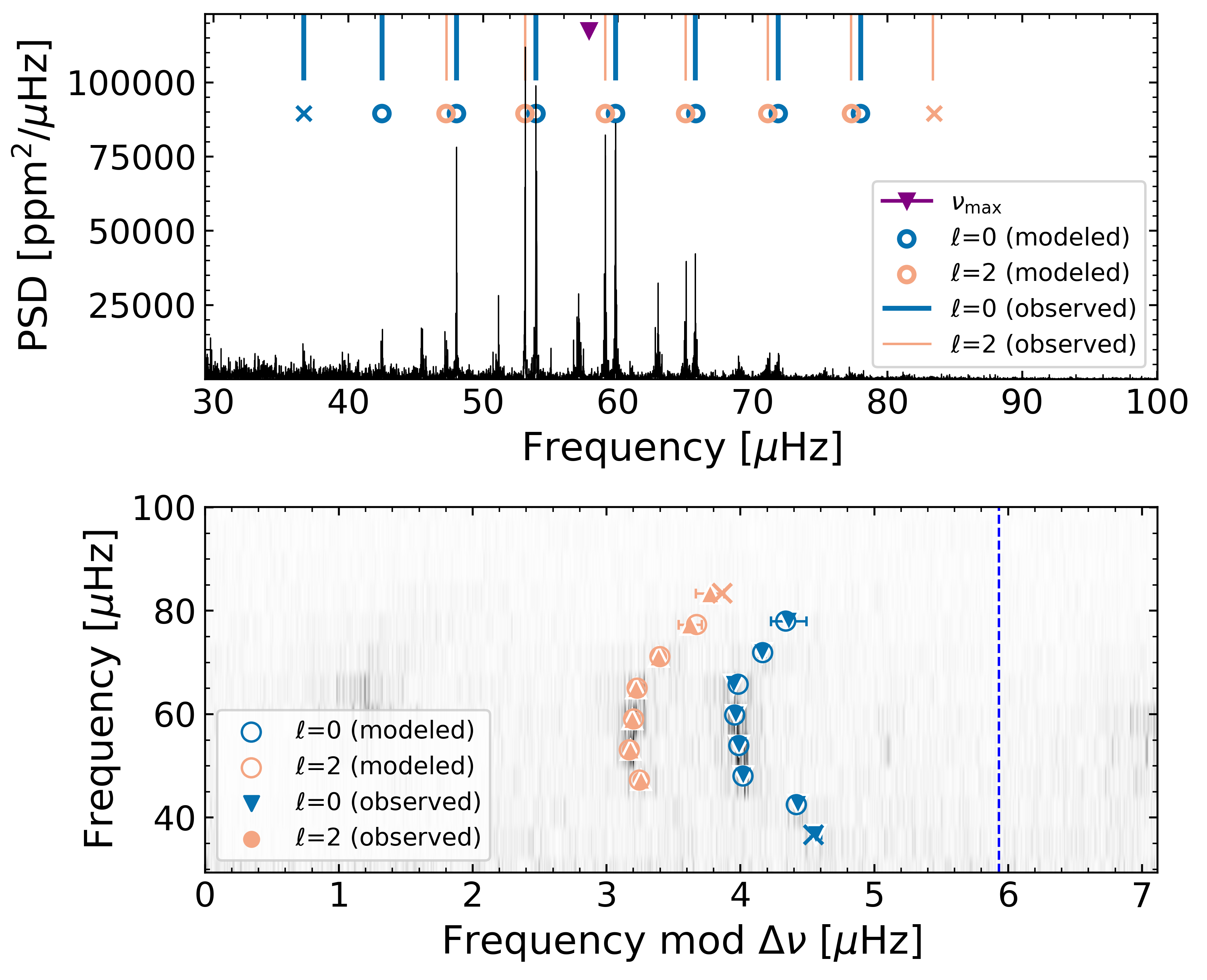}
        \caption{Top panel shows power spectral density (PSD) centered on \num for KIC\,9025498 with the modes from \cite{Kallinger2019} marked as vertical colored lines according to the legend. Those excluded from modeling due to criteria described in Sec.\,\ref{sec:modeling} are marked with a cross. Bottom panel displays the échelle diagram for the same target. The colors and symbols are the same as used in Fig.\,\ref{fig:echellefernando}.\looseness=-1}
        \label{fig:echelle}%
\end{figure}

As a next step, we applied our individual mode frequency modeling to the main sample, as defined in Sect.\,\ref{sec:mainsample}. An example échelle diagram for a star within the main sample is shown in Fig.\,\ref{fig:echelle}, displaying the modes from \cite{Kallinger2019} and those from the best-fitting model. 
After a two-step quality check of the modeling results, we removed 35 stars from the sample. First, we inspected the $\chi^2$-age plane for each star. Convergence was deemed absent when the $\chi^2$ distribution showed no well-defined minimum but instead remained approximately flat over most or all of the explored parameter space, with the determined minimum $\chi^2$ mostly being a singular outlier in the distribution. Second, for all stars flagged in the first step, we visually compared the $\ell=0$ and $\ell=2$ mode frequencies of the best-fitting model to the observed frequencies in the échelle diagram. In all 35 flagged cases, the best-fitting model failed to reproduce the observed mode pattern, confirming the absence of a physical solution within our grid. Almost all of those targets (95\,\%) did not show a Gaussian-like convergence in the Monte-Carlo simulations (Sect.\,\ref{sec:optimization}), further supporting the case for their exclusion. We investigated whether the non-converging stars occupied specific regions of the HRD or showed particular ranges of \num and \dnu, but found no clear systematic pattern. We suggest that the non-convergence is most likely related to two effects. First, these stars show on average a larger discrepancy between the seismic surface gravity of the best-fitting model and the spectroscopic surface gravity than the rest of the sample, suggesting an inherent tension between the seismic and spectroscopic constraints that prevents a well-defined $\chi^2$ minimum from forming. Second, some of these stars may be red clump stars misidentified as RGB stars, as both populations overlap significantly in the HRD and cannot be reliably distinguished without measuring the period spacing of dipole mixed modes \citep{Bedding2011}. Since our grid exclusively covers the RGB, stopping before its tip, such misidentified stars would not be reproducible, given the spectroscopic and seismic constraints, by any model in the grid \citep[e.g.,][]{campante2019,Schimak2026}. Eventually, we also removed 10 stars that had modeled masses closer than 5\,\% in mass and/or metallicity to the edges of our modeling grid, in order to minimize edge effects. The numeric results for the main sample are given in Appendix\,\ref{app:results}. 
\begin{figure*}[t]  
    \centering
    \includegraphics[width=0.33\textwidth,height=65mm]{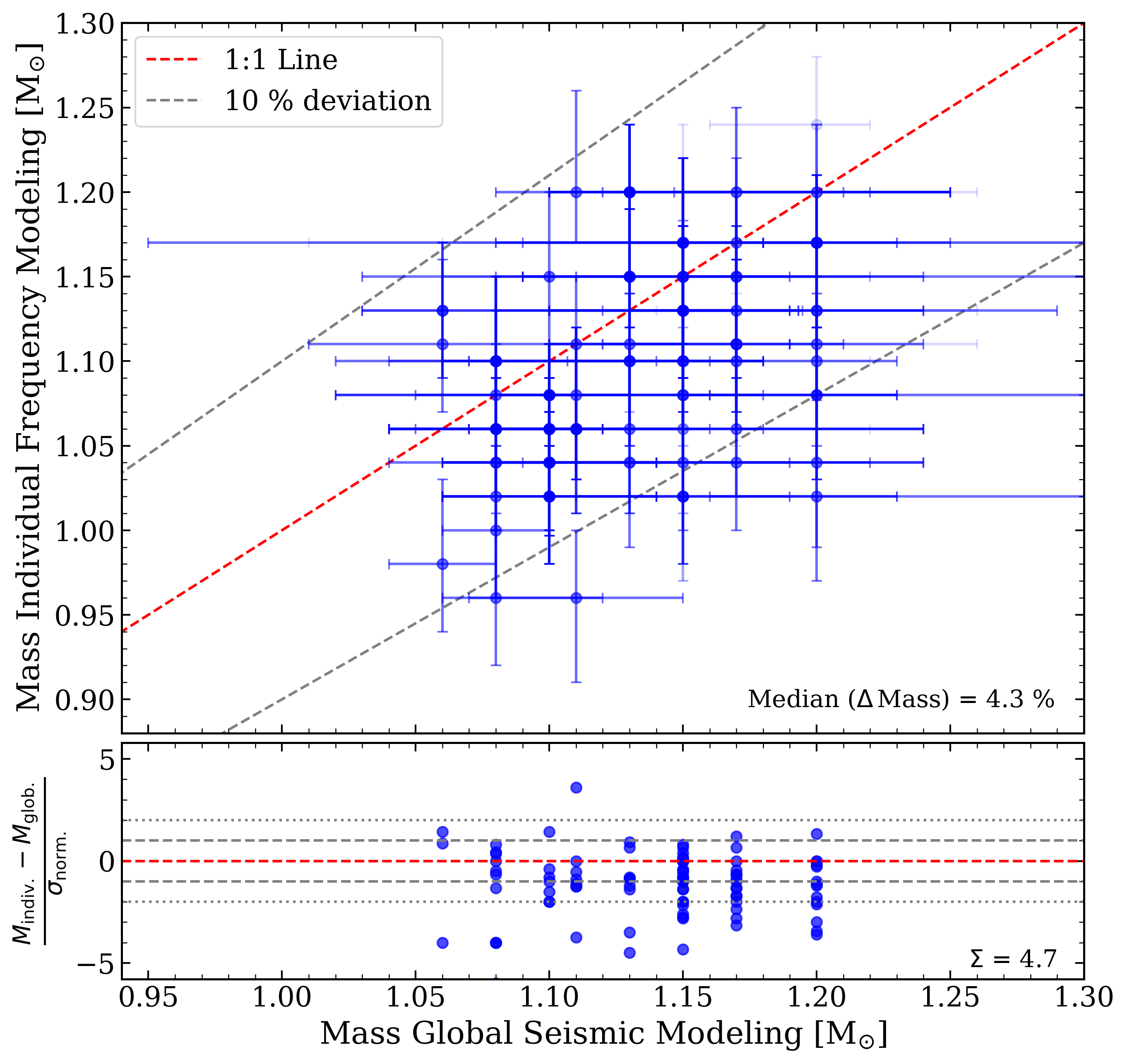}%
    \includegraphics[width=0.33\textwidth,height=65mm]{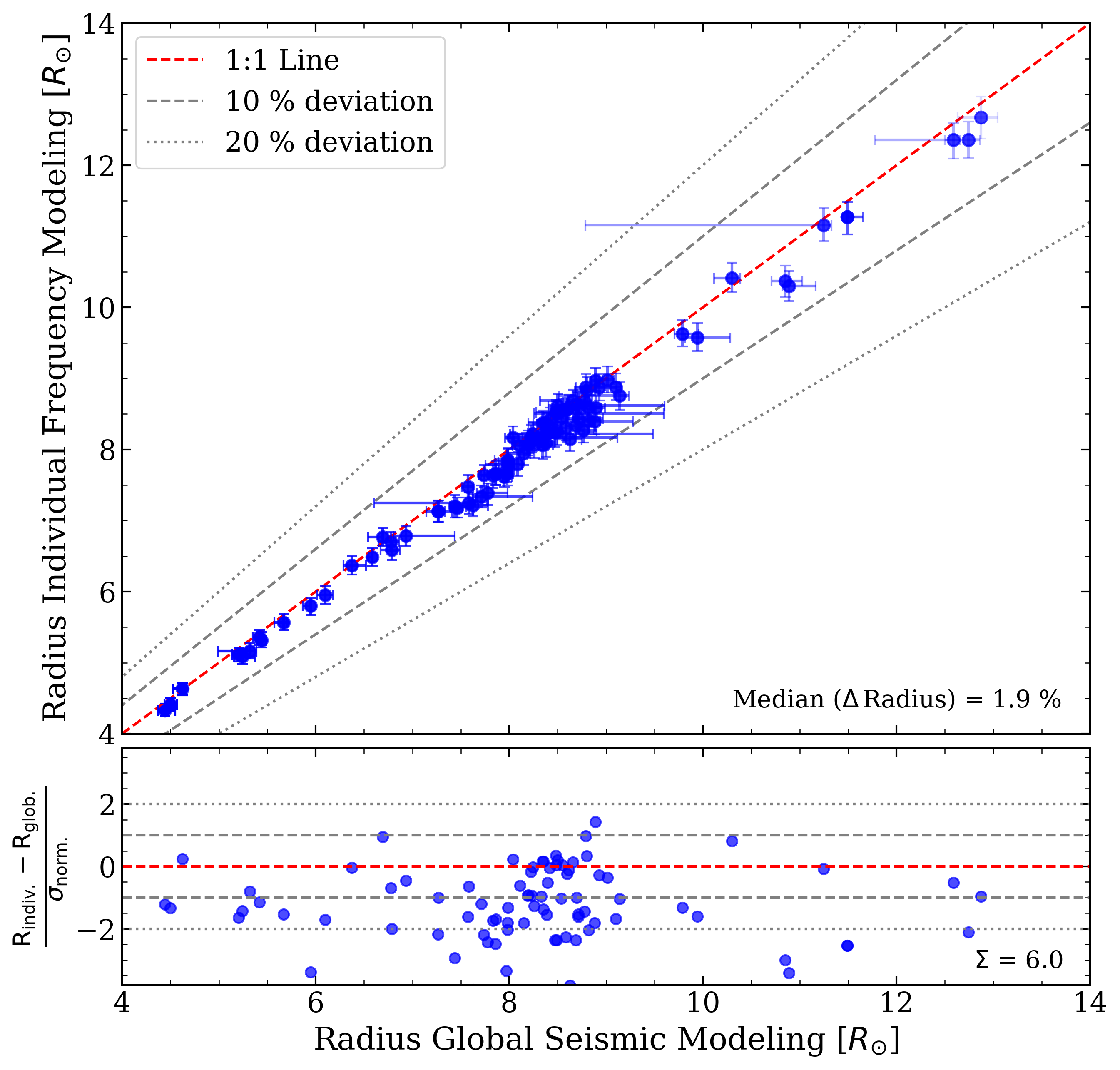}%
    \includegraphics[width=0.33\textwidth,height=65mm]{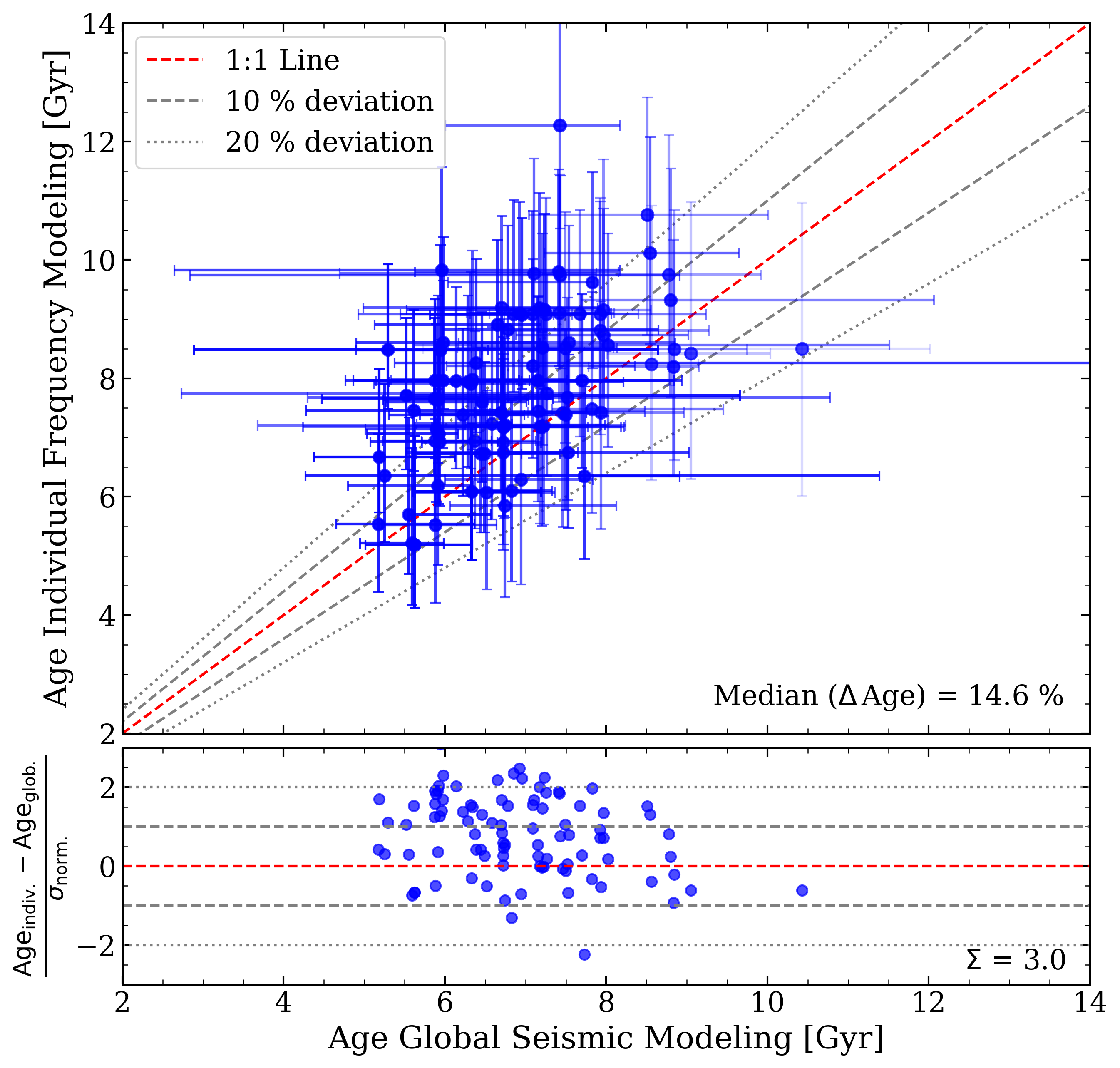}
    \caption{Comparison of mass (left top panel), radius (middle top panel), and age (right top panel) derived from global seismic mass-constrained modeling (horizontal axis) and individual mode frequency modeling (vertical axis). The respective bottom panels show the residuals, in terms of $\sigma$, between the modeled parameters from the two methods, as a function of the parameter derived from global seismic mass-constrained modeling. The dashed line and dotted grey lines represent 1$\sigma$ and 2$\sigma$ deviation, respectively.\looseness=-1}
    \label{fig:main_comparison}
\end{figure*}

We obtain a median random mass uncertainty of 3.8\,\% and 3.3\,\%, for the 16\textsuperscript{th}- and 84\textsuperscript{th}-percentile respectively. Regarding age, we obtain median random uncertainties of 12.6\,\%  and 13.8\,\% for the lower and upper uncertainties, respectively. For the derived radii, the median random uncertainties are 1.14\,\% and 1.34\,\% for the 16th- and 84th percentiles, respectively. The averages of the uncertainties are reported in Table\,\ref{tab:uncertainties}.

 As done with the validation sample, we compared the input effective temperatures with those emerging from our modeling. We report an excellent agreement between the input and output \teff ($\Sigma$\,=\,0.7). Regarding metallicity, for all stars, we obtained values well within the uncertainty of the input values. 
 For the helium abundance, the output distribution is rather flat across the explored range, with small peaks at the edges of the grid at $Y = 0.24$ and $Y = 0.30$, approximately 15\,\% higher than the remainder of the distribution. This flat distribution indicates that the individual frequency data do not strongly constrain the helium abundance, allowing models across the full grid range to contribute to the posterior, which in turn directly propagates into the age uncertainty. The elevated counts at the grid edges are consistent with the grid-boundary bias described by \cite{Valle2024}, who demonstrate that, independent of the fitting method, helium abundances in grid-based modeling are biased towards the edges of the explored ranges, a behavior also noted in \cite{Grossmann2025}. The mixing length parameter presents a flat distribution among the allowed values, with no value being preferred by the modeling.\looseness=-1

Comparing the results from the individual mode frequency modeling with those from global seismic mass-constrained modeling (Sect.\,\ref{sec:globresults}) yielded differences in the results for stellar masses (see left panel of Fig.\,\ref{fig:main_comparison}). This comparison clearly shows the discrete masses in our grid of models with a stepsize of 0.02 \Msun. On average, the masses from global modeling were 0.05 \Msun (4.3\,\%) higher compared to those from individual mode frequency modeling, with a $\Sigma$\,=\,4.7 sample level significance. The age comparison shows  similarly pronounced differences (see right panel of Fig.\,\ref{fig:main_comparison}), with the individual modeling yielding on average 14.6\,\% higher ages than the global modeling, with a $\Sigma$ = 3.0 significance. Regarding radii, we obtained, in median, 2 \% lower radii from individual modeling compared to our global modeling results, corresponding to a $\Sigma$\,=\,6.0 significance on a sample level. Therefore, all fundamental stellar parameters diverge with high significance between the two asteroseismic modeling approaches.

Given that we use the same stellar models and best-fit model search for both the global seismic mass-constrained modeling and the individual mode frequency modeling, this allows us to quantify the improvement of adding the frequencies of the modes, both in terms of systematic shifts and reduction in random uncertainties. The age differences between global seismic mass-constrained and individual frequency modeling can likely be attributed to additional information on core properties encoded in the individual frequencies that is not fully captured by \dnu and \num. Recently, \citet{reyes2025a} demonstrated the diagnostic power of individual oscillation modes by showing that individual frequencies provide constraints on the deepening of the convective zone in evolved stars.

 We note that our uncertainties only account for random uncertainties but do not include systematic uncertainties that may arise from the choice of input physics or the figure of merit. We further note that surface corrections in a parameterized way, as applied here and commonly in the literature \citep[e.g.][]{Fernando2016,Schimak2026}, can introduce biases into modeling results, not captured by our uncertainty estimation. As discussed in Sect.\,\ref{sec:surfcorr}, we verified that our surface corrections do not shift the corrected frequencies into non-physical ranges. However, we acknowledge that this does not preclude the existence of a systematic bias. Since the surface correction affects all frequencies in a correlated manner, its impact does not average out with the number of modes, and may therefore dominate over the random uncertainties. This represents a known limitation shared by all asteroseismic studies employing parameterized surface corrections, and a quantitative assessment of this systematic effect is beyond the scope of the present work.
 
 Finally, we note that two stars in our sample have \Gaia-based luminosities within 3\,$\sigma$ of the upper luminosity boundary of the modeling grid ($\log\,(L/L_{\odot}) = 1.8$; Sect.\,\ref{sec:modeling}). These targets, KIC\,8737047 and KIC\,9277886, exhibit larger-than-average age uncertainties and less well-defined minima in their $\chi^2$-age planes, and may therefore be subject to edge effects introduced by the proximity to the grid boundary. In particular, KIC\,9277886 displays an age uncertainty more than twice the sample average. We caution that the derived ages for these two stars should be interpreted carefully.

\subsection{Individual mode frequency modeling including luminosities}
\label{sec:lum}

We also tested how including luminosity as an input parameter of the modeling impacts the derived stellar parameters. To do so, we took the luminosities of our targets from \cite{Godoy25}, who computed luminosities for around 170,000 \Kepler targets, using the \GaiaDR parallax as an input. \cite{Godoy25} provides several different bolometric corrections using a variety of spectroscopic surveys. For consistency, we used the luminosity calculated with the bolometric corrections using spectroscopic parameters from the APOGEE survey. For further details of the bolometric correction and luminosity estimation, we refer to \cite{Godoy25}.\looseness=-1

\begin{figure}[t!]
        \centering
     \includegraphics[width=\columnwidth
]{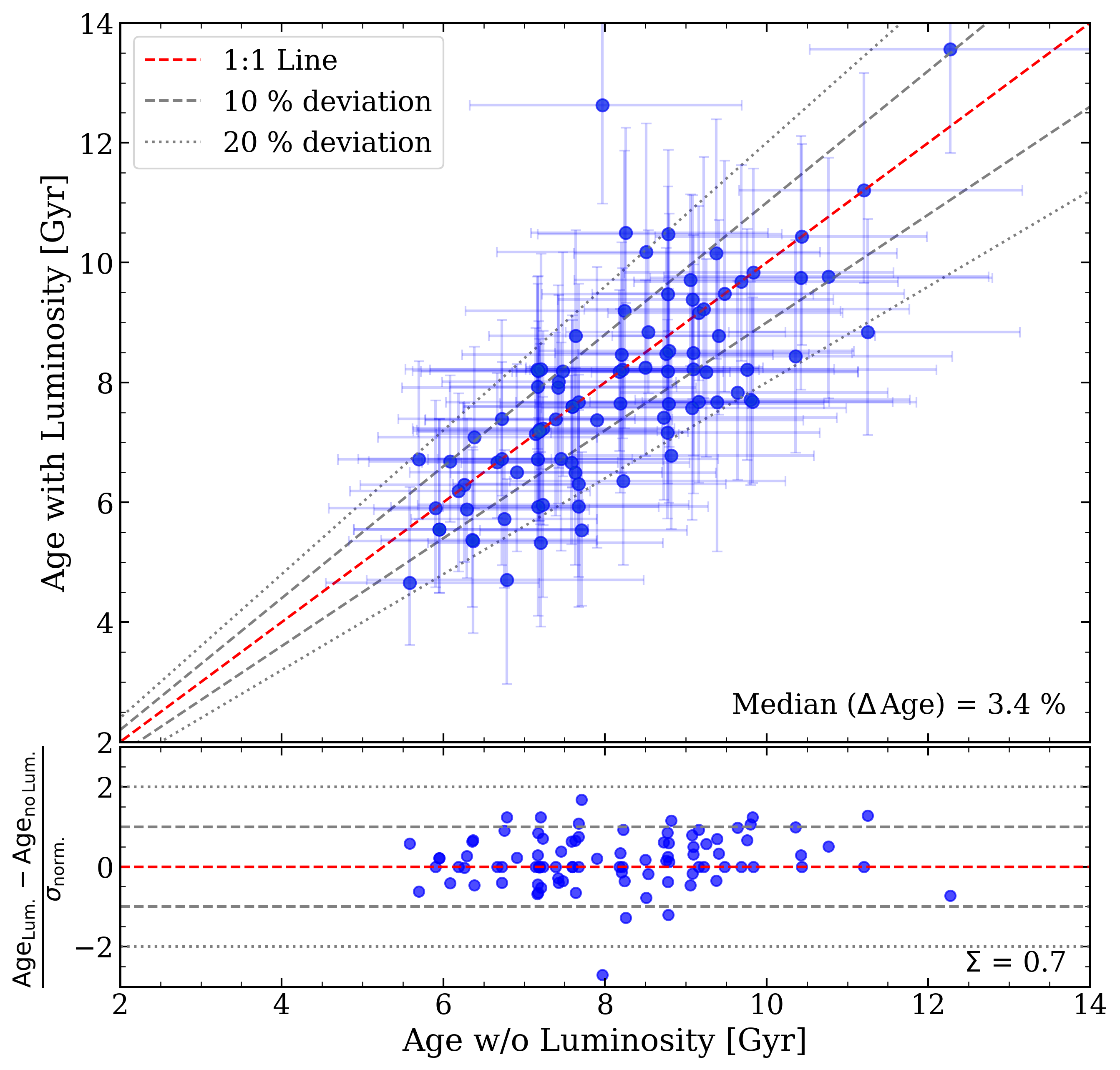}
        \caption{Top: Comparison of ages from modeling without \Gaia luminosity and with \Gaia luminosity. Bottom: Residuals, in terms of $\sigma$, between the two age estimations as a function of age without luminosity. The dashed line and dotted grey lines represent 1$\sigma$ and 2$\sigma$ deviation, respectively.}
        \label{fig:lumhist}%
\end{figure}

We recomputed the best-fitting parameters, including these luminosities as inputs (i.e., adding a $\chi^2_L$ term to Equation \ref{eq:chifull}). Figure\,\ref{fig:lumhist} shows the comparison of ages from the modeling, including luminosity as an additional constraint, with those derived without. The median difference between the two methods is 3.4\,\% ($\Sigma$\,=\,0.7), indicating no significant systematic offset on average. However, the star-by-star scatter is substantial, and for individual stars the inclusion of luminosity can noticeably shift the derived age. 
We attribute this scatter primarily to cases where the \Gaia luminosity is in tension with the seismic and spectroscopic constraints: when the luminosity implied by the parallax is inconsistent with what the oscillation frequencies and atmospheric parameters predict, the figure of merit is forced into a compromise solution that deviates from the seismic-only result. Unresolved binarity is another possible reason of such tension, as an undetected companion with comparable luminosities could inflate the apparent luminosity and consequently shifts the inferred age. Additional contributions from luminosity measurement uncertainties such as parallax zero-point offsets \citep{lindegren2021}, extinction corrections, and bolometric corrections are likely present but secondary. Regarding stellar mass, we obtain a median difference of less than 0.02\,\Msun, below the stepsize in mass in the modeling grid. We therefore conclude that \Gaia luminosity constraints introduce no significant population-level bias ($\Sigma$\,=\,0.7), but can affect individual stellar ages at the 15\% level.\looseness=-1

We note that the grid-like pattern visible in Fig.\,\ref{fig:lumhist} is likely a visual artifact of the error bar display. Since mass is a discrete grid parameter, stars with similar properties tend to have error bars of similar discrete lengths, whose extremities naturally align when over plotted. Inspection of the numerical age values confirms that the underlying ages are distinct for each star and do not exhibit an actual discretization.
\vspace{-0.5em}
\subsection{Literature comparison}
\begin{figure}[t!]
        \centering
     \includegraphics[width=\columnwidth
]{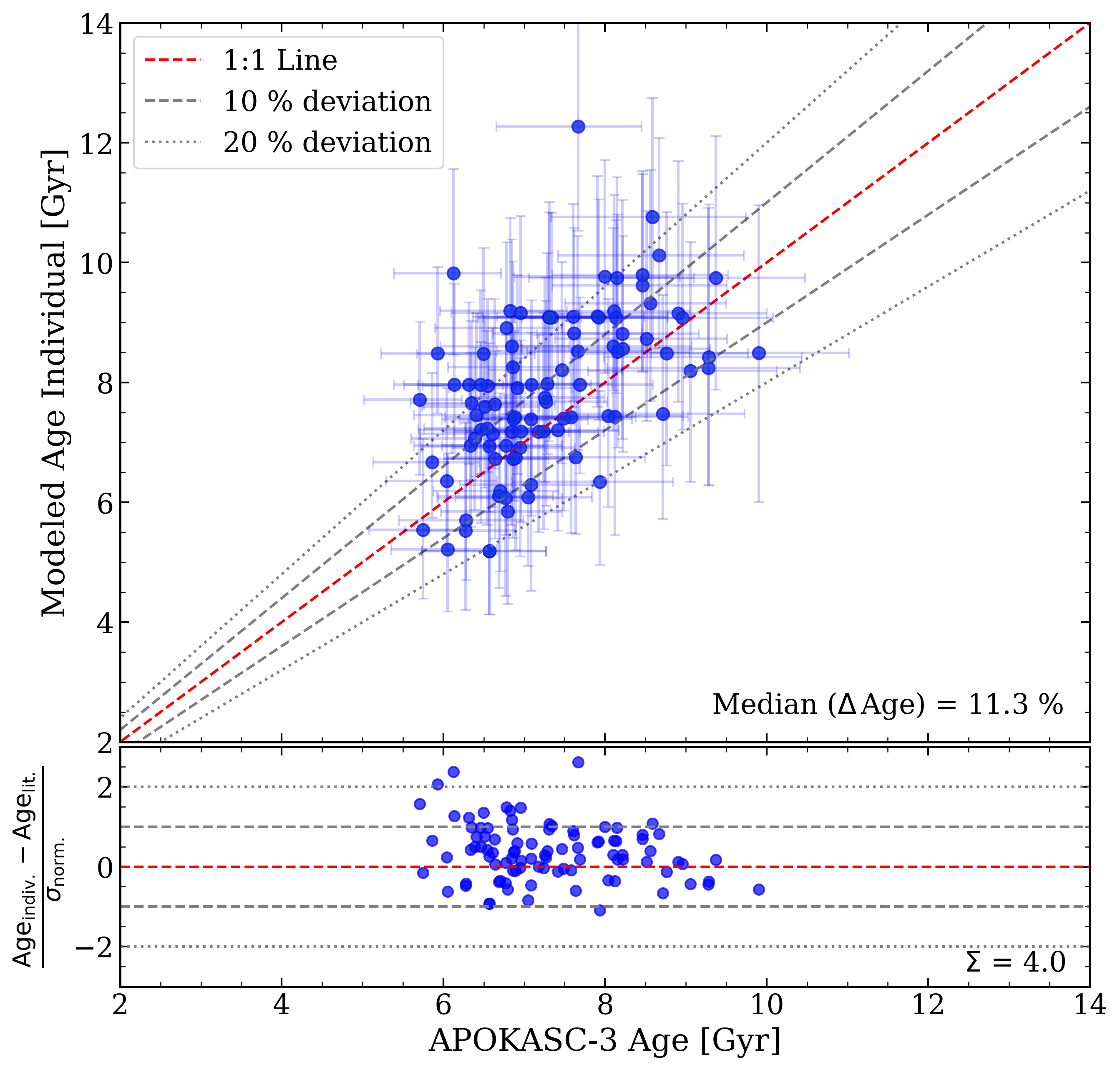}
        \caption{Top panel comparison of ages from APOKASC-3 (horizontal axis) and ages from individual mode frequency modeling (vertical axis). Bottom panel shows the residuals, in terms of $\sigma$, between age from APOKASC-3 and individual mode frequency modeling as a function of the APOKASC-3 age. The dashed line and dotted grey lines represent 1$\sigma$ and 2$\sigma$ deviation, respectively.}
        \label{fig:apoindiv}%
\end{figure}

\vspace{-0.5em}
\label{sec:lit}
Comparing our results from modeling, including individual mode-frequencies, without the luminosity constraint, with the literature values from \cite{APOKASC3} yielded a similar result to Section\,\ref{sec:globresults}. Figure\,\ref{fig:apoindiv} shows a comparison of the ages from APOKASC-3 with those obtained by our individual mode frequency modeling. On average, following stellar evolution theory, the ages retrieved from individual mode frequency modeling are higher, and the corresponding masses are lower, compared to the values in APOKASC-3. The median differences in mass, radius, and age were found to be 4.3\,\%, 2.2\%, and 11.3\,\%, respectively. The sample level significances for those differences are $\Sigma$\,=\,5.7 (mass), $\Sigma$\,=\,6.9 (radius) and $\Sigma$\,=\,4.0 (age), hence a significant offset of stellar parameters compared to APOKASC-3.

\begin{figure}[t!]
        \centering
     \includegraphics[width=\columnwidth
]{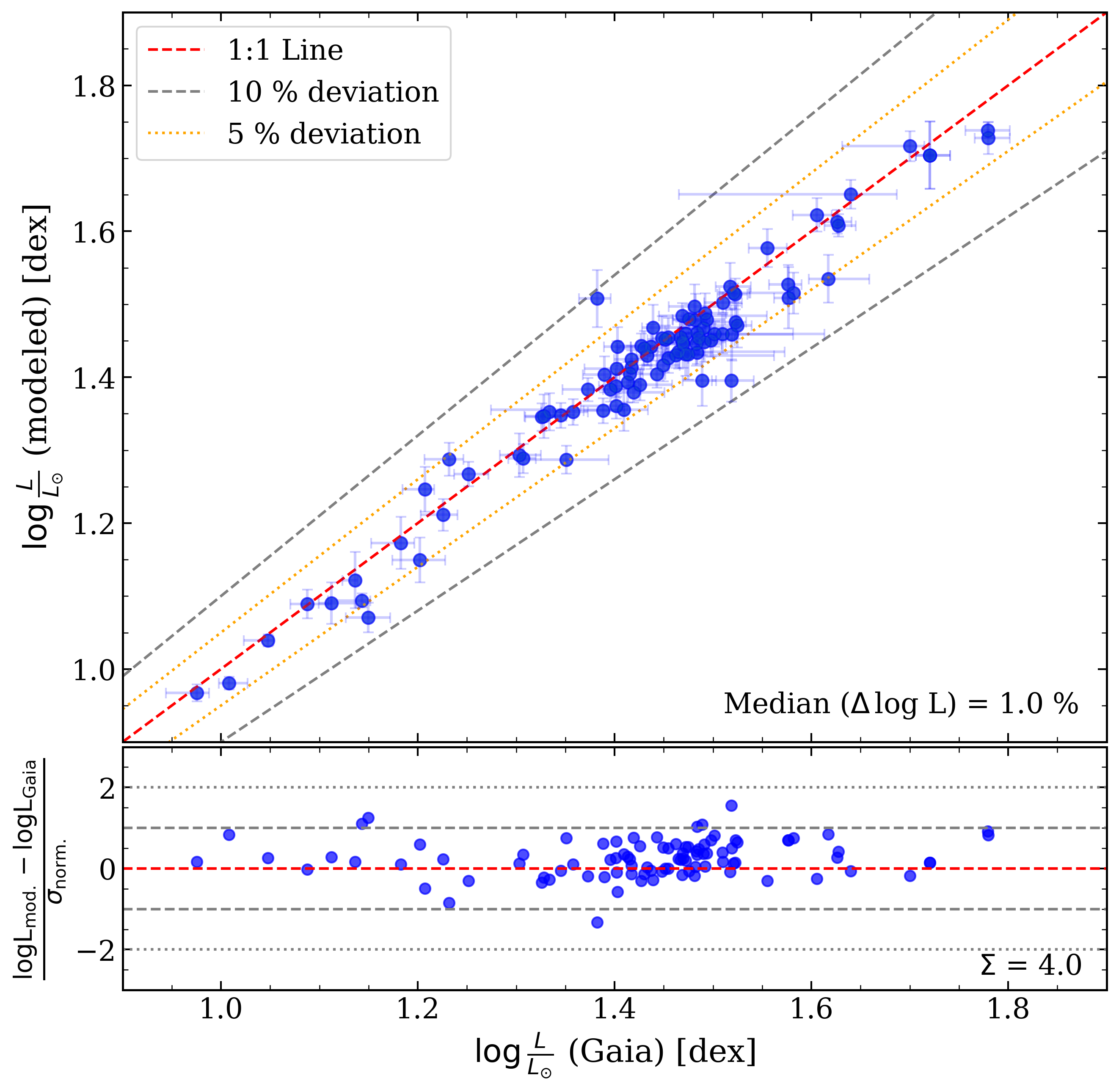}
        \caption{Top panel displays the comparison of luminosities from modeling with corrected luminosities from \GaiaDR. Bottom panel shows the residuals, in terms of $\sigma$, between the modeled and \Gaia luminosity as a function of \Gaia luminosity.}
        \label{fig:lumcomp}%
\end{figure}

As an independent benchmark of our modeling results, we compared the luminosities obtained with our individual mode frequency modeling with the corrected \Gaia luminosities as described in Sect.\,\ref{sec:lum}. For this comparison, we take the models without the luminosity constraint. The only common quantity between our modeling and the corrected luminosities from \GaiaDR is the effective temperature from APOKASC-3. This temperature is an input quantity in our modeling and is used for the bolometric correction to obtain the luminosities from \Gaia. However, this should not significantly impact the comparison, as it only influences the luminosity evaluation on a small scale \citep{Godoy25}.
As seen in Fig.\,\ref{fig:lumcomp}, the corrected luminosities from \GaiaDR and the values of our best fit modeling show a very good agreement. Except for four outliers, all targets fall within 5\,\% of the one-to-one line. All of those outliers also have a higher than average deviation in the age comparison in the previous section (global vs individual seismic modeling). The median deviation of the logarithmic luminosities obtained by individual mode frequency modeling from the corrected \Gaia luminosities 1.0\,\% ($\Sigma$\,=\,4.0), while in the global-seismic modeling case, the median difference is 1.9\%. We additionally compared our radii to the fundamental Gaia radii provided in APOKASC-3 \citep{APOKASC3}, finding a median deviation of 1.5\,\% ($\Sigma$\,=\,2.7), consistent with no systematic zero-point offset and further supporting the validity of our modeling results. The agreement in luminosities, and radii, also demonstrates that the asteroseismic parallaxes, if obtained with careful modeling, provide a reliability comparable to those acquired by \Gaia, especially at larger distances \citep{Huber2017}.\looseness=-1

\begin{figure}
        \centering
     \includegraphics[width=\columnwidth
]{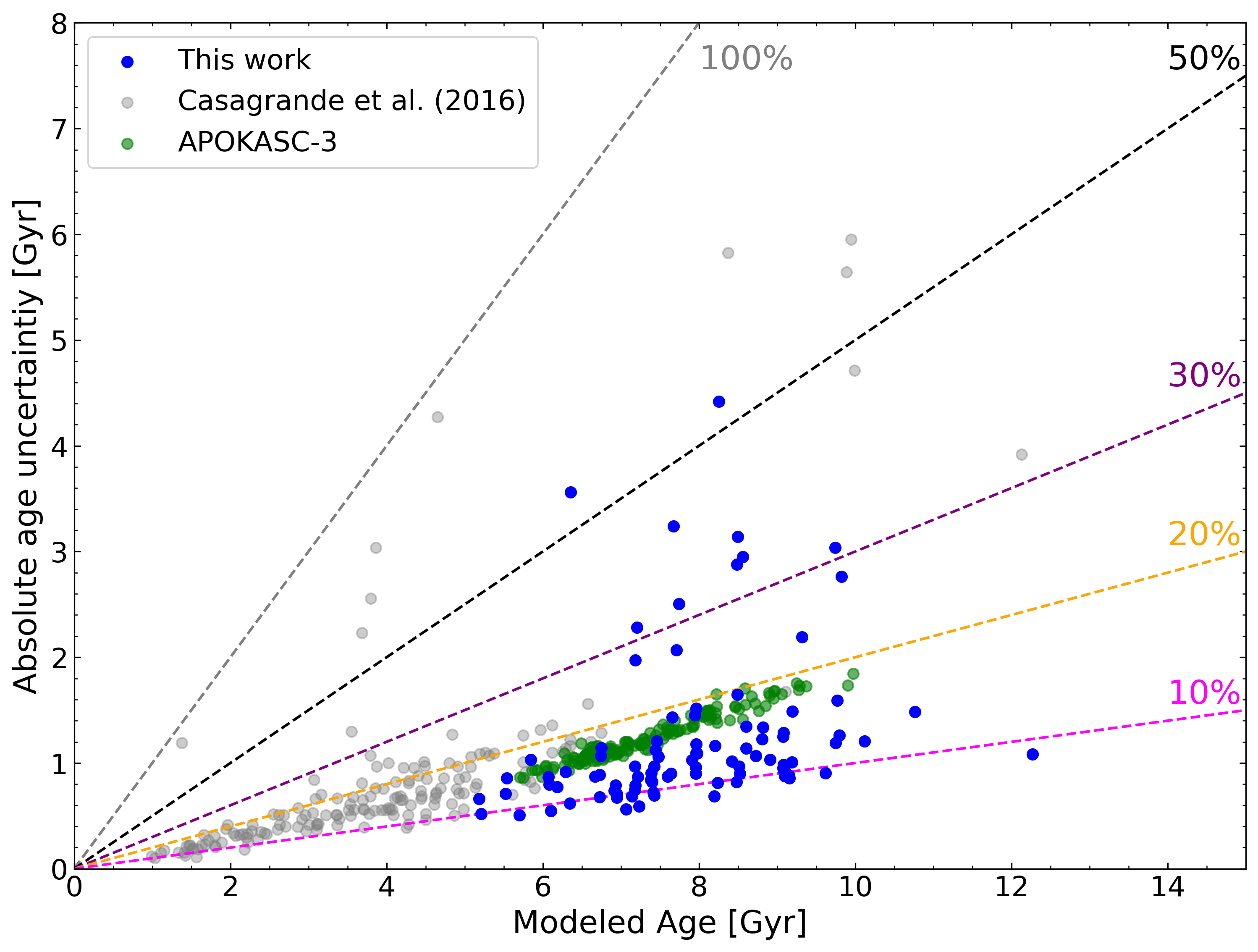}
        \caption{Absolute age uncertainty over age for the sample of RGB stars in  \cite{Cassagrande2016} (grey), the ages for our sample coming from individual mode frequency modeling (blue) and, from APOKASC3 (green). The colored dashed lines represent the relative age uncertainties.}
        \label{fig:casagrande}%
\end{figure}
To put our results into context, we compared them with age uncertainties derived by \cite{Cassagrande2016}, who used a combination of isochrones and global asteroseismic parameter fitting for age determination. Similarly to Fig.\,2 of \cite{Cassagrande2016}, we show in Fig.\,\ref{fig:casagrande} the absolute age uncertainties as a function of age, for the individual mode frequency modeling results of our sample, the APOKASC-3 values for the same stars, and all RGB stars in \cite{Cassagrande2016}. To account for our asymmetric age uncertainties, we computed the square root of the sum of the quadratic upper and lower uncertainties. The majority of our sample (90\,\%) falls in the range of relative age uncertainties below 20\,\%, mostly close to 10\,\%, for an age range from around 5\,Gyr to 11\,Gyr. We conclude that seismic ages based on individual mode frequencies modeling are typically more precise than those derived using only global seismic parameters. Furthermore, Fig.\,\ref{fig:casagrande} demonstrates that our results extend the mass range to lower (solar-like) masses, and hence older ages, compared to those RGB stars studied in \cite{Cassagrande2016}. As the scope of this work focused on stars around solar mass and metallicity, no stars with younger or older ages are part of the selected sample.

\begin{figure}[t!]
        \centering
     \includegraphics[width=\columnwidth
]{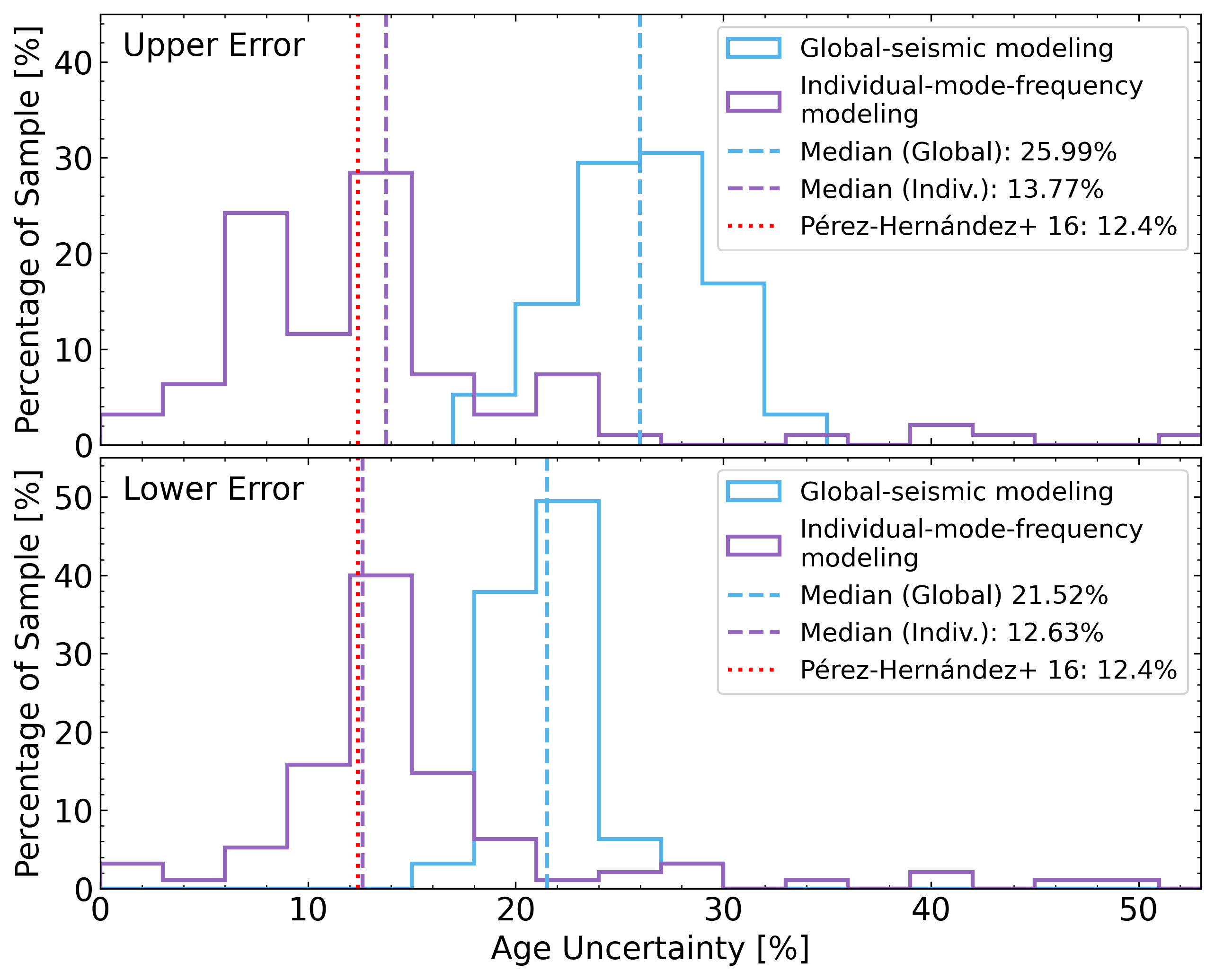}
        \caption{Histogram of relative random age uncertainties for upper error (upper panel) and lower error (lower panel), for global seismic mass-constrained mass-constrained modeling (light-blue) and individual mode frequency modeling (purple). The medians of the distributions are shown as vertical dashed lines, the median age uncertainty from \cite{Fernando2016} is shown as a dotted red line. }
        \label{fig:uncertaintyhist}%
\end{figure}

As depicted in Fig.\,\ref{fig:uncertaintyhist}, comparing the median uncertainties from individual mode frequency modeling and global seismic mass-constrained modeling demonstrates that adding the individual frequencies as constraints to the figure of merit improves the random age uncertainties by a factor of $\approx2$. The random age uncertainties from individual mode frequency modeling are also close to the median uncertainty from individual mode frequency modeling in \cite{Fernando2016}, given as 12.4\,\%.\looseness=-1

\vspace{-3mm}
\section{Conclusions}
\label{sec:conclusion}
In this work, we have presented an ensemble modeling of red-giant stars, including their individual mode frequencies, by combining asteroseismic data from \Kepler and spectroscopic data from APOGEE with a modeling grid of stellar evolution and oscillation models using MESA and GYRE. 
First, we validated our method using 16 RGB stars with previous individual mode frequency modeling from the literature. Subsequently, we applied the method to a sample of 94 RGB stars with close to solar properties, which lack current individual mode frequency modeling, and thereby derived precise stellar masses and ages for this sample.\looseness=-1

We have compared the results from our individual mode frequency modeling with those of a custom global modeling approach, as well as with values from the literature.
This work presents one of the largest samples of RGB stars to date that has comprehensive individual mode frequency modeling, providing precise stellar masses and ages, with random uncertainties yielding approximately 3\,\% and 13\,\% for mass and age, respectively. We have found that ages from global seismic mass-constrained modeling are on average 14\,\% systematically younger compared to those derived from individual mode frequency modeling, while masses are around 0.05 \Msun systematically higher when obtained with global seismic mass-constrained modeling. Consistently, radii obtained from individual modeling are on average 2\% lower than those from global seismic mass-constrained modeling. These systematic shifts in mass and age between the methods, that are statistically significant, are in line with results from recent studies for selected individual stars \citep[e.g.][]{Larsen2025,Lindsay2025}. Most of these studies focused on selected metal-poor stars. However, this work demonstrates that such a discrepancy in age between the methods is also present in stars with properties close to the Sun. We conclude that fitting individual oscillation modes introduces constraints on the internal stellar structure that go beyond those provided by the global seismic mass-constrained parameters, which likely accounts for the systematic age difference found between the two approaches. Furthermore, we note that using different approaches for global seismic modeling can introduce systematic shifts in the age estimation of up to 0.4 Gyr.\looseness=-1

Regarding the random age precision, we have found that adding individual frequencies in the figure of merit can reduce the relative age, mass, and radius uncertainties by a factor of about 1.5 to 2. We conclude that by including individual modes additional information is added in the modeling that can constrain ages significantly more as compared to global seismic modeling. The obtained random uncertainties in age are in the regime of $\approx10\,\%$, across an age range of 5-11\,Gyr. This degree of precision for a wide age range is especially important for galactic archaeology studies \citep[e.g.][]{Gallart2019} and, if acquired for a larger sample of stars across the parameter space, can be a significant contribution to constrain the assembly history and (chemical) evolution of our Galaxy. 

Especially important for galactic archaeology are halo stars. Those metal-poor stars outside the Milky Way's disks belong to the oldest populations known and can be tracers of the assembly history of our Galaxy. However, due to their poor signal-to-noise ratio in asteroseismic observations, those targets are often poorly characterized. To date, only very few Halo stars have precise parameters from individual mode frequency modeling \citep[e.g.][]{chaplin2020, lundkvist2025a}. However, those works \citep[e.g.,][]{Huber2024,Lindsay2025} have demonstrated that classical age and mass estimation methods - such as the asteroseismic scaling relations \citep[for details see review of][]{hekker2020} or global seismic modeling - overestimate mass and underestimate age of those targets significantly, on even a larger scale than found for stars close to the Sun in this work. Therefore, modeling a larger sample of Halo stars would provide us with more robust masses and ages for this old and distant stellar population.\looseness=-1

Future space missions such as ESA PLAnetary Transits and Oscillations of stars \citep[PLATO;][]{Rauer2014,Rauer2024} Mission or NASA Nancy Grace Roman Space Telescope \citep{Roman} will contribute significantly to the fraction of low-mass stars with detected solar-like oscillations \citep{RomanRGSeismology2023,Goupil2024,Weiss2025,scvPIC_Zwintz2026}, and in particular enhance the sky coverage significantly compared to \Kepler. Asteroseismic data from those future missions, combined with high-quality spectroscopy from surveys like APOGEE \citep{APOGEE},  Large Sky Area Multi-Object Fibre Spectroscopic Telescope \citep[LAMOST;][]{Lamost} or  William Herschell Telescope Enhanced Area Velocity Explorer \citep[WEAVE;][]{WEAVE}, 4-metre Multi-Object Spectroscopic Telescope \citep[4MOST;][]{4MOST} and astrometric data from the upcoming Data Release 4 from \Gaia \citep{GaiaDR4proceed} will enable us to study the history and evolution of our Galaxy with unprecedented precision.\looseness=-1

\section*{Data availability}
Table~\ref{tab:resultsmain} is available at the CDS via anonymous ftp to \url{cdsarc.u-strasbg.fr} ([???]) or via \url{http://cdsweb.u-strasbg.fr/cgi-bin/qcat?J/A+A/[???]} and at in electronic form on Zenodo (\url{https://zenodo.org/records/[???]}).
\begin{acknowledgements}
We thank the anonymous referee for the constructive feedback, that helped to improve this manuscript.
The authors thank the people behind the ESA \Gaia and NASA \Kepler mission.  The authors wish to acknowledge the contribution of the IAC  High-Performance Computing support team and hardware facilities to the results of this research.

The project that gave rise to these results received the support of a fellowship from ”la Caixa” Foundation (ID 100010434). The fellowship code is LCF/BQ/DI23/11990068. This work was supported by the {International Space Science Institute (ISSI) in Bern (project\,24-629).} This work is part of grant CEX2025-001609-S, awarded to the Instituto de Astrofísica de Canarias under the Severo Ochoa Centre of Excellence program and funded by MICIU/AEI/10.13039/501100011033.
PGB acknowledges support by the Spanish Ministry of Science, Innovation and Universities (MCIN) with the \textit{Ram{\'o}n\,y\,Cajal} fellowship (RYC-2021-033137-I, MRR4032204). DHG, PGB and DGR acknowledge support from MCIN for the project \textit{PLAtoSOnG} from its grant PID2023-146453NB-I00) and DHG, SM, DGR and RAG acknowledge similar support from grant PID2023-149439NB-C41. 
DGR acknowledges support from the \textit{Juan\,de\,la\,Cierva} program under contract JDC2022-049054-I.  DBP, RAG, acknowledge the support from the  PLATO Centre National D'{\'{E}}tudes Spatiales grant. LSS acknowledges support from the Australian Research Council through Laureate Fellowship FL220100117.\\

This work has made use of data from the European Space Agency (ESA) mission
\href{https://www.cosmos.esa.int/gaia}{\Gaia}, processed by the \href{https://www.cosmos.esa.int/web/gaia/dpac/consortium}{\Gaia
Data Processing and Analysis Consortium (DPAC)}. Funding for the DPAC
has been provided by national institutions, in particular the institutions
participating in the \Gaia Multilateral Agreement. This paper includes data collected by the \textit{Kepler} mission and obtained from the MAST data archive at the Space Telescope Science Institute (STScI). Funding for the Kepler mission is provided by the NASA Science Mission Directorate. STScI is operated by the Association of Universities for Research in Astronomy, Inc., under NASA contract NAS 5–26555.

\end{acknowledgements}

\bibliographystyle{aa} 
\bibliography{bib-refs.bib} 

\appendix
\section{Sample selection}
\begin{figure*}
        \centering
        \includegraphics[width=\textwidth
]{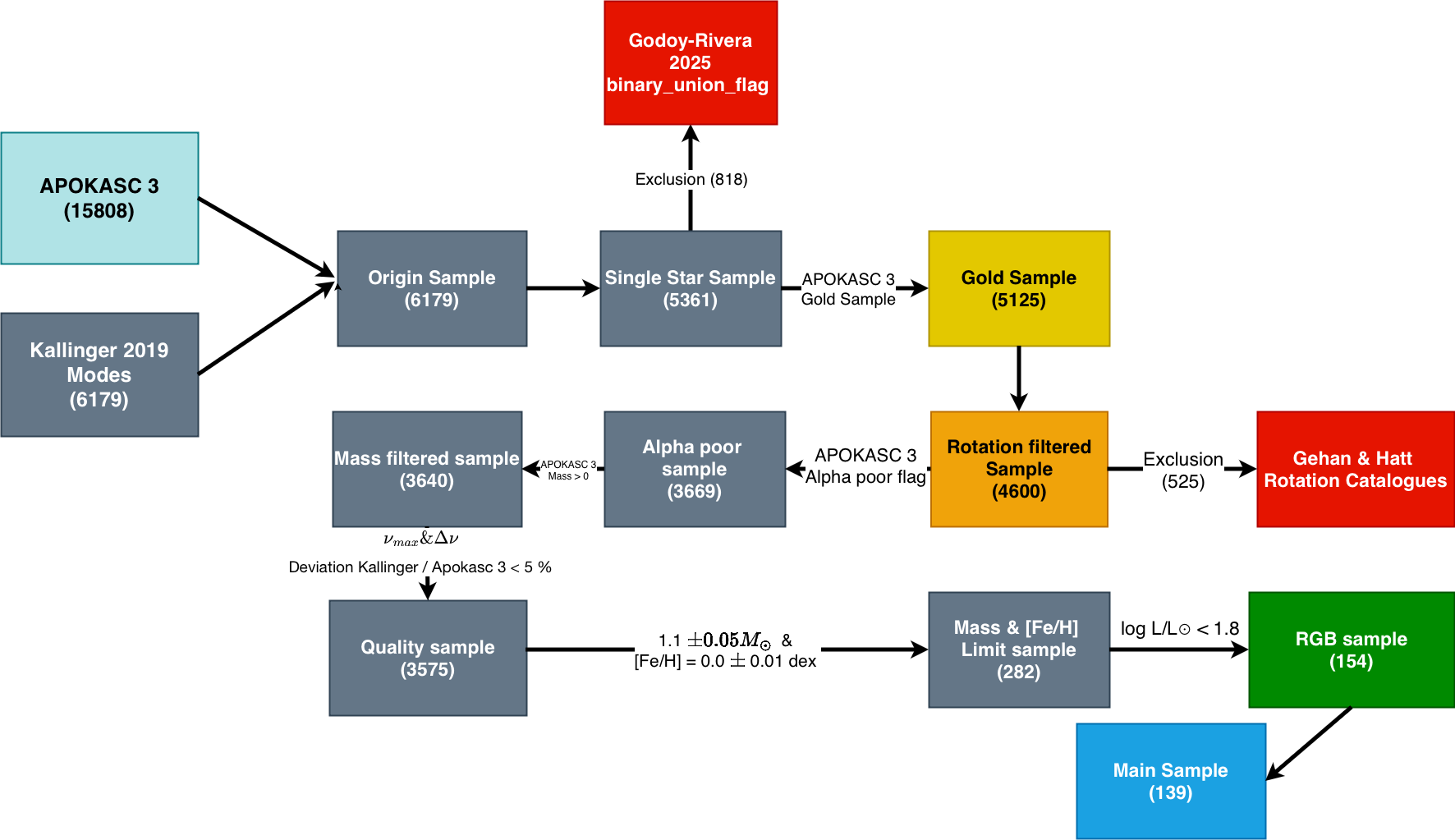}
        \caption{Visual depiction of the sample selection for the main sample as described in Sect. \ref{sec:mainsample}.}
        \label{fig:sample}%
\end{figure*}

Figure\,\ref{fig:sample} provides a visual depiction of the sample selection procedure for our main sample of 139 stars described in Sect.\,\ref{sec:sample}.

\section{Mode Calculation} \label{app:gyre}
\subsection{MESA inlist} \label{app:MESA}
In Listing\,\ref{list:MESA}, we show an example MESA inlist used for the computation of the model grid. For concise display, some lines were broken with the Fortran line break \&, the variables over the grid are marked accordingly.

\begin{lstlisting}[language=Fortran,
    caption = MESA example inlist., 
    label= list:MESA,frame=Tb]

&star_job

      history_columns_file = &
      history_columns_list_path
      show_log_description_at_start = .false.
      create_pre_main_sequence_model = .true.

      save_model_when_terminate = .true.
      save_model_filename = 'RGB.mod'
      required_termination_code_string &
      = 'log_L_upper_limit'
      color_num_files = 2
      color_file_names(1) = 'lcb98cor.dat'
      color_num_colors(1) = 11
      color_num_files = 2
      color_file_names(2) = colorfilepath
      color_num_colors(2) = 17
      pgstar_flag = .false.
      pause_before_terminate = .false.

      relax_to_this_tau_factor = 0.001
      relax_tau_factor = .true.
/ ! end of star_job namelist

&eos

/ ! end of eos namelist

&kap

      kap_file_prefix = 'gs98'
      use_Type2_opacities = .true.

/ ! end of kap namelist

&controls

      log_L_upper_limit = 1.8d0  !
      mixing_length_alpha = 1.5 ! variable
      initial_mass = 1.39 ! variable
      initial_z = 0.006 ! variable
      initial_y = 0.27 ! variable
      num_trace_history_values = 2
      trace_history_value_name(1) = &
      'rel_E_err'
      trace_history_value_name(2) = &
      'log_rel_run_E_err'

      cool_wind_full_on_T = 9.99d9
      hot_wind_full_on_T = 1d10
      cool_wind_RGB_scheme = 'Reimers'
      RGB_to_AGB_wind_switch = 1d-4
      Reimers_scaling_factor = 0.2

      varcontrol_target = 1d-3
      delta_lgL_He_limit = 0.01d0
      time_delta_coeff = 0.5

     !mesh
      mesh_delta_coeff = 1
      mesh_min_dr_div_cs = -1
      max_allowed_nz = 30000
      use_other_mesh_delta_coeff_factor = &
      .false.

     !output
      photo_interval = -1
      profile_interval = 5
      history_interval = 1
      terminal_interval = 100
      write_header_frequency = 50

      atm_T_tau_opacity = 'varying'
      atm_T_tau_relation = 'Eddington'

      !overshooting
      overshoot_scheme(1) = 'exponential'
      overshoot_zone_type(1) = 'any'
      overshoot_zone_loc(1) = 'any'
      overshoot_bdy_loc(1) = 'any'
      overshoot_f0(1) = 0.005
      overshoot_f(1) = 0.02
     
      max_num_profile_models = 10000
      max_model_number = 50000

/ ! end of controls namelist

&pgstar

/ ! end of pgstar namelist
\end{lstlisting}

\subsection{GYRE example inlist} \label{app:GYRE}
Listing\,\ref{list:GYRE} shows an example inlist for computing individual model frequencies with GYRE. The parameters in curly brackets following a \$ represent the variables that change over the course of stellar evolution as described in Sect.\,\ref{sec:modeling}.
\begin{lstlisting}[language=Fortran,
    caption = GYRE example inlist, 
    breaklines=true,
    label= list:GYRE,frame=Tb]
&constants
/

&model
  model_type = 'EVOL'  ! Obtain stellar structure from an evolutionary model
  file = 'profile${counter}.data.FGONG'    ! File name of the evolutionary model
  file_format = 'FGONG' ! File format of the evolutionary model
/

&mode
  l = 0 ! Harmonic degree
  tag = 'radial'
/

&mode
  l = 2
  tag = 'non_radial'
/

&osc
  outer_bound = 'ISOTHERMAL' ! uses an isothermal atmosphere prescription
/

&rot
/

&num
  diff_scheme = 'COLLOC_GL4' ! 4th-order collocation scheme for difference equations
/

&scan
  grid_type = 'LINEAR'
  freq_min = ${LOWER}      ! Minimum frequency to scan from
  freq_max = ${UPPER}      ! Maximum frequency to scan to
  n_freq = 100        ! Number of frequency points in scan
  freq_min_units = 'UHZ'
  freq_max_units = 'UHZ'
  tag_list = 'radial'
/

&scan
  grid_type = 'LINEAR' 
  freq_min = ${LOWER}     ! Minimum frequency to scan from
  freq_max = ${UPPER}      ! Maximum frequency to scan to
  n_freq = 400        ! Number of frequency points in scan
  freq_min_units = 'UHZ'
  freq_max_units = 'UHZ'
  tag_list = 'non_radial'
/

&grid
  w_osc = 2 ! Oscillatory region weight parameter
  w_exp = 1  ! Exponential region weight parameter
  w_ctr = 4 ! Central region weight parameter
/


&ad_output
  summary_file = 'summary${counter}.h5'                         ! File name for summary file
  summary_item_list = 'l,n_pg,n_p,n_g,freq,freq_units,E_norm, M_star, R_star, L_star, zeta, E' ! Items to appear in summary file
  freq_units = 'UHZ'                   	      ! Units of freq output items
/

&nad_output
/
\end{lstlisting}

\section{Stellar Parameters from individual mode frequency modeling} 
Table\,\ref{app:results} displays the KIC identifiers, fundamental stellar parameters from modeling for all targets of the main sample, both for the case of individual mode frequency modeling and global-seismic modeling. This table is also provided in a machine-readable format via CDS.

\onecolumn 
{\setlength{\tabcolsep}{3pt}
\renewcommand{\arraystretch}{1.27}
\begin{longtable}{l | c c c c c || c c c c c}
\caption{Stellar parameters from individual mode frequency and global seismic mass-constrained modeling for the main sample.}\label{tab:resultsmain} \\
\hline\hline
KIC & Age$_{\rm indiv.}$ & Mass$_{\rm indiv.}$ & Radius$_{\rm indiv.}$ & $T_{\rm eff, indiv.}$ & $\log g_{\rm indiv.}$ & Age$_{\rm glob.}$ & Mass$_{\rm glob.}$ & Radius$_{\rm glob.}$ & $T_{\rm eff, glob.}$ & $\log g_{\rm glob.}$ \\
 & [Gyr] &  [M$_\odot$] &  [R$_\odot$] & [K] & [dex] & [Gyr] &  [M$_\odot$] &  [R$_\odot$] & [K] & [dex]\\
\hline
2016978 & $6.93^{+0.72}_{-0.86}$ & $1.13^{+0.07}_{-0.05}$ & $7.13^{+0.14}_{-0.13}$ & $4690^{+37}_{-39}$ & $2.783^{+0.006}_{-0.005}$ & $5.93^{+1.72}_{-1.09}$ & $1.17^{+0.05}_{-0.04}$ & $7.27^{+0.15}_{-0.15}$ & $4606^{+131}_{-125}$ & $2.783^{+0.003}_{-0.004}$ \\
2305987 & $7.39^{+0.01}_{-1.81}$ & $1.08^{+0.18}_{-0.02}$ & $8.69^{+0.13}_{-0.34}$ & $4625^{+46}_{-45}$ & $2.625^{+0.019}_{-0.039}$ & $7.50^{+1.44}_{-1.61}$ & $1.15^{+0.05}_{-0.04}$ & $8.66^{+0.15}_{-0.19}$ & $4515^{+114}_{-138}$ & $2.625^{+0.004}_{-0.004}$ \\
2424955 & $5.54^{+1.19}_{-0.53}$ & $1.11^{+0.04}_{-0.02}$ & $8.59^{+0.10}_{-0.03}$ & $4678^{+39}_{-43}$ & $2.612^{+0.005}_{-0.001}$ & $5.18^{+1.52}_{-1.15}$ & $1.20^{+0.04}_{-0.05}$ & $8.89^{+0.17}_{-0.18}$ & $4619^{+128}_{-126}$ & $2.617^{+0.004}_{-0.004}$ \\
2446327 & $9.20^{+1.27}_{-1.71}$ & $1.06^{+0.07}_{-0.02}$ & $8.30^{+0.13}_{-0.12}$ & $4586^{+42}_{-42}$ & $2.628^{+0.008}_{-0.007}$ & $6.70^{+1.55}_{-1.81}$ & $1.15^{+0.07}_{-0.04}$ & $8.59^{+0.19}_{-0.18}$ & $4575^{+108}_{-140}$ & $2.633^{+0.004}_{-0.004}$ \\
2580691 & $9.32^{+3.27}_{-1.11}$ & $1.06^{+0.07}_{-0.04}$ & $8.06^{+0.19}_{-0.11}$ & $4586^{+30}_{-50}$ & $2.648^{+0.008}_{-0.004}$ & $8.80^{+2.23}_{-1.63}$ & $1.11^{+0.04}_{-0.05}$ & $8.26^{+0.17}_{-0.18}$ & $4506^{+99}_{-134}$ & $2.647^{+0.004}_{-0.004}$ \\
3100505 & $8.24^{+0.58}_{-1.04}$ & $1.11^{+0.02}_{-0.05}$ & $10.41^{+0.08}_{-0.19}$ & $4560^{+43}_{-38}$ & $2.443^{+0.003}_{-0.006}$ & $8.56^{+2.67}_{-1.96}$ & $1.06^{+0.05}_{-0.04}$ & $10.30^{+0.22}_{-0.19}$ & $4444^{+123}_{-124}$ & $2.435^{+0.004}_{-0.004}$ \\
3218930 & $6.75^{+1.50}_{-0.79}$ & $1.15^{+0.07}_{-0.09}$ & $8.87^{+0.18}_{-0.25}$ & $4587^{+37}_{-30}$ & $2.600^{+0.006}_{-0.011}$ & $7.53^{+1.65}_{-1.28}$ & $1.15^{+0.05}_{-0.05}$ & $8.93^{+0.19}_{-0.18}$ & $4475^{+129}_{-151}$ & $2.591^{+0.004}_{-0.004}$ \\
3247334 & $7.44^{+1.32}_{-0.92}$ & $1.08^{+0.03}_{-0.06}$ & $6.70^{+0.08}_{-0.13}$ & $4696^{+52}_{-42}$ & $2.814^{+0.004}_{-0.005}$ & $7.16^{+1.95}_{-1.53}$ & $1.08^{+0.05}_{-0.04}$ & $6.78^{+0.13}_{-0.14}$ & $4670^{+141}_{-137}$ & $2.810^{+0.004}_{-0.003}$ \\
3325179 & $5.85^{+1.38}_{-0.68}$ & $1.15^{+0.07}_{-0.02}$ & $8.22^{+0.16}_{-0.06}$ & $4641^{+54}_{-34}$ & $2.667^{+0.006}_{-0.003}$ & $6.74^{+1.80}_{-1.55}$ & $1.17^{+0.05}_{-0.04}$ & $8.33^{+0.20}_{-0.16}$ & $4557^{+109}_{-142}$ & $2.663^{+0.004}_{-0.004}$ \\
3444374 & $9.09^{+0.73}_{-1.12}$ & $1.10^{+0.03}_{-0.04}$ & $8.23^{+0.07}_{-0.13}$ & $4584^{+35}_{-30}$ & $2.648^{+0.004}_{-0.005}$ & $7.67^{+1.75}_{-2.08}$ & $1.15^{+0.05}_{-0.04}$ & $8.39^{+0.18}_{-0.19}$ & $4537^{+85}_{-133}$ & $2.651^{+0.004}_{-0.004}$ \\
3529399 & $8.42^{+0.99}_{-1.04}$ & $1.04^{+0.02}_{-0.02}$ & $5.32^{+0.05}_{-0.01}$ & $4693^{+56}_{-55}$ & $3.002^{+0.002}_{-0.001}$ & $9.05^{+2.55}_{-2.13}$ & $1.08^{+0.05}_{-0.04}$ & $5.44^{+0.11}_{-0.10}$ & $4624^{+137}_{-136}$ & $2.999^{+0.004}_{-0.004}$ \\
3735447 & $8.81^{+1.35}_{-1.11}$ & $1.11^{+0.02}_{-0.03}$ & $7.48^{+0.06}_{-0.07}$ & $4603^{+47}_{-30}$ & $2.733^{+0.002}_{-0.004}$ & $7.93^{+2.24}_{-1.76}$ & $1.13^{+0.04}_{-0.05}$ & $7.58^{+0.17}_{-0.15}$ & $4548^{+113}_{-146}$ & $2.727^{+0.004}_{-0.003}$ \\
3750891 & $6.10^{+0.51}_{-0.59}$ & $1.20^{+0.06}_{-0.05}$ & $8.60^{+0.17}_{-0.12}$ & $4596^{+30}_{-45}$ & $2.645^{+0.004}_{-0.004}$ & $6.82^{+1.53}_{-1.54}$ & $1.20^{+0.04}_{-0.07}$ & $8.61^{+0.17}_{-0.18}$ & $4524^{+92}_{-137}$ & $2.640^{+0.004}_{-0.004}$ \\
3855126 & $6.95^{+0.70}_{-0.65}$ & $1.13^{+0.04}_{-0.03}$ & $8.24^{+0.11}_{-0.09}$ & $4662^{+39}_{-38}$ & $2.658^{+0.004}_{-0.004}$ & $5.88^{+1.56}_{-1.36}$ & $1.20^{+0.04}_{-0.05}$ & $8.47^{+0.15}_{-0.20}$ & $4594^{+107}_{-144}$ & $2.657^{+0.004}_{-0.003}$ \\
3937022 & $9.11^{+0.69}_{-1.15}$ & $1.06^{+0.02}_{-0.06}$ & $7.94^{+0.07}_{-0.16}$ & $4642^{+45}_{-45}$ & $2.659^{+0.003}_{-0.006}$ & $7.42^{+2.34}_{-1.51}$ & $1.11^{+0.04}_{-0.05}$ & $8.15^{+0.17}_{-0.16}$ & $4578^{+114}_{-140}$ & $2.658^{+0.004}_{-0.003}$ \\
4249357 & $7.96^{+1.15}_{-1.21}$ & $1.11^{+0.02}_{-0.07}$ & $8.88^{+0.06}_{-0.20}$ & $4636^{+43}_{-47}$ & $2.583^{+0.004}_{-0.007}$ & $5.98^{+1.69}_{-1.15}$ & $1.17^{+0.05}_{-0.04}$ & $9.10^{+0.19}_{-0.18}$ & $4578^{+103}_{-148}$ & $2.587^{+0.004}_{-0.003}$ \\
4634108 & $6.29^{+0.89}_{-0.94}$ & $1.17^{+0.03}_{-0.07}$ & $8.53^{+0.05}_{-0.20}$ & $4644^{+30}_{-60}$ & $2.644^{+0.003}_{-0.007}$ & $6.94^{+1.62}_{-1.77}$ & $1.15^{+0.05}_{-0.05}$ & $8.48^{+0.17}_{-0.17}$ & $4581^{+109}_{-137}$ & $2.637^{+0.004}_{-0.004}$ \\
4814689 & $8.91^{+0.55}_{-1.52}$ & $1.11^{+0.04}_{-0.05}$ & $6.59^{+0.08}_{-0.11}$ & $4675^{+40}_{-39}$ & $2.845^{+0.004}_{-0.007}$ & $6.65^{+1.43}_{-1.43}$ & $1.17^{+0.05}_{-0.04}$ & $6.78^{+0.13}_{-0.14}$ & $4589^{+143}_{-121}$ & $2.845^{+0.004}_{-0.004}$ \\
4830883 & $10.12^{+1.10}_{-1.31}$ & $1.08^{+0.03}_{-0.02}$ & $8.60^{+0.09}_{-0.05}$ & $4549^{+44}_{-38}$ & $2.600^{+0.003}_{-0.003}$ & $8.55^{+1.96}_{-1.54}$ & $1.10^{+0.03}_{-0.04}$ & $8.71^{+0.20}_{-0.17}$ & $4479^{+136}_{-125}$ & $2.594^{+0.005}_{-0.003}$ \\
4915387 & $7.96^{+1.15}_{-0.78}$ & $1.06^{+0.04}_{-0.04}$ & $8.59^{+0.12}_{-0.10}$ & $4610^{+30}_{-40}$ & $2.592^{+0.004}_{-0.004}$ & $7.70^{+1.46}_{-1.48}$ & $1.11^{+0.06}_{-0.03}$ & $8.82^{+0.19}_{-0.17}$ & $4523^{+129}_{-158}$ & $2.593^{+0.004}_{-0.004}$ \\
4937784 & $7.68^{+3.25}_{-3.22}$ & $1.15^{+0.21}_{-0.07}$ & $8.18^{+0.37}_{-0.09}$ & $4611^{+52}_{-47}$ & $2.661^{+0.044}_{-0.013}$ & $7.52^{+1.69}_{-1.74}$ & $1.15^{+0.05}_{-0.04}$ & $8.22^{+0.17}_{-0.17}$ & $4544^{+94}_{-158}$ & $2.670^{+0.004}_{-0.004}$ \\
5005173 & $8.49^{+0.91}_{-2.38}$ & $1.10^{+0.03}_{-0.06}$ & $8.37^{+0.24}_{-0.15}$ & $4588^{+35}_{-36}$ & $2.631^{+0.005}_{-0.062}$ & $8.84^{+2.36}_{-1.87}$ & $1.08^{+0.05}_{-0.04}$ & $8.34^{+0.17}_{-0.18}$ & $4522^{+108}_{-137}$ & $2.629^{+0.004}_{-0.004}$ \\
5175621 & $5.53^{+0.76}_{-0.66}$ & $1.20^{+0.04}_{-0.05}$ & $8.83^{+0.09}_{-0.12}$ & $4652^{+44}_{-42}$ & $2.622^{+0.003}_{-0.005}$ & $5.88^{+1.79}_{-1.32}$ & $1.17^{+0.05}_{-0.04}$ & $8.80^{+0.19}_{-0.16}$ & $4616^{+121}_{-151}$ & $2.614^{+0.004}_{-0.004}$ \\
5183325 & $9.08^{+0.41}_{-2.16}$ & $1.10^{+0.03}_{-0.08}$ & $7.78^{+0.07}_{-0.23}$ & $4596^{+47}_{-30}$ & $2.694^{+0.004}_{-0.006}$ & $7.09^{+1.74}_{-1.67}$ & $1.15^{+0.05}_{-0.04}$ & $7.99^{+0.17}_{-0.16}$ & $4566^{+102}_{-128}$ & $2.693^{+0.004}_{-0.003}$ \\
5358184 & $9.08^{+0.49}_{-1.41}$ & $1.13^{+0.02}_{-0.05}$ & $7.86^{+0.01}_{-0.13}$ & $4603^{+30}_{-30}$ & $2.698^{+0.001}_{-0.005}$ & $6.85^{+1.93}_{-1.32}$ & $1.15^{+0.07}_{-0.04}$ & $7.98^{+0.17}_{-0.17}$ & $4555^{+99}_{-133}$ & $2.696^{+0.004}_{-0.003}$ \\
5544447 & $7.91^{+0.86}_{-1.20}$ & $1.13^{+0.02}_{-0.07}$ & $8.12^{+0.06}_{-0.18}$ & $4619^{+50}_{-30}$ & $2.669^{+0.003}_{-0.007}$ & $6.32^{+1.88}_{-1.44}$ & $1.15^{+0.05}_{-0.04}$ & $8.23^{+0.16}_{-0.16}$ & $4593^{+105}_{-134}$ & $2.667^{+0.004}_{-0.004}$ \\
5610212 & $6.67^{+0.94}_{-0.81}$ & $1.17^{+0.02}_{-0.02}$ & $6.48^{+0.01}_{-0.04}$ & $4777^{+30}_{-53}$ & $2.881^{+0.001}_{-0.002}$ & $5.18^{+1.49}_{-0.93}$ & $1.20^{+0.04}_{-0.05}$ & $6.58^{+0.12}_{-0.12}$ & $4712^{+134}_{-145}$ & $2.879^{+0.004}_{-0.004}$ \\
5688694 & $6.73^{+0.93}_{-0.84}$ & $1.15^{+0.02}_{-0.04}$ & $8.98^{+0.06}_{-0.12}$ & $4641^{+51}_{-40}$ & $2.590^{+0.002}_{-0.005}$ & $6.49^{+1.55}_{-1.33}$ & $1.13^{+0.04}_{-0.03}$ & $9.01^{+0.19}_{-0.18}$ & $4604^{+104}_{-138}$ & $2.582^{+0.003}_{-0.004}$ \\
5869293 & $7.96^{+1.79}_{-1.24}$ & $1.06^{+0.09}_{-0.04}$ & $8.43^{+0.25}_{-0.11}$ & $4593^{+30}_{-37}$ & $2.608^{+0.011}_{-0.006}$ & $7.15^{+1.40}_{-1.61}$ & $1.15^{+0.05}_{-0.05}$ & $8.71^{+0.18}_{-0.17}$ & $4555^{+111}_{-126}$ & $2.614^{+0.004}_{-0.004}$ \\
6041223 & $7.21^{+1.24}_{-0.49}$ & $1.17^{+0.03}_{-0.02}$ & $8.63^{+0.05}_{-0.07}$ & $4587^{+41}_{-30}$ & $2.633^{+0.002}_{-0.001}$ & $6.74^{+1.46}_{-1.56}$ & $1.20^{+0.04}_{-0.05}$ & $8.69^{+0.17}_{-0.18}$ & $4559^{+85}_{-159}$ & $2.631^{+0.004}_{-0.004}$ \\
6345456 & $9.77^{+0.77}_{-2.41}$ & $1.02^{+0.04}_{-0.04}$ & $5.16^{+0.07}_{-0.33}$ & $4783^{+54}_{-68}$ & $3.019^{+0.074}_{-0.006}$ & $7.11^{+1.94}_{-1.61}$ & $1.10^{+0.03}_{-0.04}$ & $5.32^{+0.12}_{-0.11}$ & $4746^{+142}_{-136}$ & $3.023^{+0.004}_{-0.004}$ \\
6429995 & $9.15^{+0.59}_{-1.17}$ & $0.98^{+0.02}_{-0.02}$ & $7.79^{+0.05}_{-0.07}$ & $4654^{+31}_{-61}$ & $2.646^{+0.005}_{-0.004}$ & $7.97^{+2.54}_{-1.48}$ & $1.06^{+0.05}_{-0.04}$ & $8.08^{+0.17}_{-0.16}$ & $4613^{+112}_{-153}$ & $2.647^{+0.003}_{-0.004}$ \\
6504840 & $7.95^{+1.93}_{-0.98}$ & $1.02^{+0.04}_{-0.02}$ & $8.41^{+0.12}_{-0.06}$ & $4677^{+36}_{-48}$ & $2.594^{+0.005}_{-0.003}$ & $6.29^{+1.45}_{-1.44}$ & $1.15^{+0.05}_{-0.04}$ & $8.82^{+0.16}_{-0.17}$ & $4603^{+110}_{-148}$ & $2.608^{+0.004}_{-0.004}$ \\
6751420 & $8.73^{+1.05}_{-1.08}$ & $1.15^{+0.02}_{-0.07}$ & $6.77^{+0.01}_{-0.15}$ & $4660^{+55}_{-30}$ & $2.837^{+0.005}_{-0.003}$ & $7.97^{+2.14}_{-1.37}$ & $1.10^{+0.05}_{-0.04}$ & $6.69^{+0.13}_{-0.12}$ & $4599^{+112}_{-131}$ & $2.828^{+0.004}_{-0.004}$ \\
6768319 & $7.21^{+1.02}_{-3.55}$ & $1.20^{+0.12}_{-0.03}$ & $8.62^{+1.10}_{-0.10}$ & $4587^{+37}_{-30}$ & $2.642^{+0.002}_{-0.060}$ & $7.22^{+1.93}_{-1.68}$ & $1.13^{+0.04}_{-0.03}$ & $8.50^{+0.16}_{-0.16}$ & $4569^{+96}_{-126}$ & $2.631^{+0.003}_{-0.004}$ \\
6775098 & $7.23^{+0.25}_{-0.94}$ & $1.20^{+0.02}_{-0.05}$ & $8.36^{+0.03}_{-0.11}$ & $4608^{+46}_{-30}$ & $2.669^{+0.003}_{-0.005}$ & $6.58^{+1.63}_{-1.62}$ & $1.20^{+0.04}_{-0.05}$ & $8.40^{+0.15}_{-0.17}$ & $4563^{+75}_{-147}$ & $2.664^{+0.004}_{-0.004}$ \\
6777074 & $6.95^{+0.75}_{-0.66}$ & $1.20^{+0.02}_{-0.05}$ & $8.36^{+0.01}_{-0.10}$ & $4631^{+38}_{-38}$ & $2.668^{+0.001}_{-0.005}$ & $6.37^{+1.85}_{-1.49}$ & $1.17^{+0.05}_{-0.06}$ & $8.35^{+0.17}_{-0.17}$ & $4577^{+107}_{-130}$ & $2.660^{+0.003}_{-0.004}$ \\
6844373 & $7.42^{+0.54}_{-0.91}$ & $1.06^{+0.02}_{-0.02}$ & $8.06^{+0.06}_{-0.05}$ & $4666^{+38}_{-39}$ & $2.646^{+0.004}_{-0.002}$ & $7.46^{+1.71}_{-1.93}$ & $1.13^{+0.04}_{-0.05}$ & $8.34^{+0.17}_{-0.20}$ & $4565^{+108}_{-137}$ & $2.647^{+0.004}_{-0.004}$ \\
6853127 & $7.07^{+0.24}_{-0.89}$ & $1.13^{+0.09}_{-0.05}$ & $4.33^{+0.11}_{-0.07}$ & $4818^{+67}_{-42}$ & $3.218^{+0.009}_{-0.006}$ & $5.92^{+1.52}_{-1.19}$ & $1.20^{+0.04}_{-0.05}$ & $4.44^{+0.09}_{-0.08}$ & $4783^{+124}_{-162}$ & $3.216^{+0.004}_{-0.004}$ \\
6870779 & $8.52^{+0.60}_{-1.33}$ & $1.10^{+0.03}_{-0.02}$ & $8.41^{+0.08}_{-0.05}$ & $4606^{+50}_{-50}$ & $2.626^{+0.005}_{-0.004}$ & $7.49^{+2.29}_{-1.60}$ & $1.08^{+0.05}_{-0.02}$ & $8.42^{+0.18}_{-0.16}$ & $4594^{+106}_{-139}$ & $2.622^{+0.004}_{-0.004}$ \\
6890257 & $9.82^{+2.21}_{-3.32}$ & $1.02^{+0.13}_{-0.04}$ & $8.40^{+0.40}_{-0.13}$ & $4617^{+38}_{-30}$ & $2.595^{+0.007}_{-0.025}$ & $5.96^{+1.74}_{-1.36}$ & $1.20^{+0.04}_{-0.05}$ & $8.88^{+0.18}_{-0.19}$ & $4563^{+109}_{-142}$ & $2.618^{+0.003}_{-0.004}$ \\
6937257 & $9.08^{+0.76}_{-1.14}$ & $1.04^{+0.04}_{-0.02}$ & $7.65^{+0.11}_{-0.06}$ & $4654^{+39}_{-47}$ & $2.686^{+0.004}_{-0.003}$ & $6.96^{+1.63}_{-1.26}$ & $1.10^{+0.03}_{-0.04}$ & $7.86^{+0.15}_{-0.15}$ & $4631^{+120}_{-150}$ & $2.686^{+0.004}_{-0.004}$ \\
7187059 & $7.18^{+1.04}_{-0.89}$ & $1.15^{+0.02}_{-0.04}$ & $8.45^{+0.06}_{-0.11}$ & $4610^{+38}_{-44}$ & $2.643^{+0.002}_{-0.005}$ & $6.73^{+1.82}_{-1.55}$ & $1.15^{+0.07}_{-0.04}$ & $8.54^{+0.17}_{-0.19}$ & $4575^{+104}_{-146}$ & $2.638^{+0.004}_{-0.004}$ \\
7345128 & $9.19^{+0.72}_{-1.29}$ & $1.04^{+0.04}_{-0.04}$ & $5.95^{+0.08}_{-0.08}$ & $4745^{+34}_{-39}$ & $2.903^{+0.005}_{-0.005}$ & $7.17^{+1.94}_{-1.32}$ & $1.10^{+0.03}_{-0.04}$ & $6.10^{+0.13}_{-0.12}$ & $4702^{+131}_{-152}$ & $2.905^{+0.003}_{-0.004}$ \\
7446648 & $7.60^{+0.55}_{-1.19}$ & $1.11^{+0.09}_{-0.05}$ & $5.08^{+0.14}_{-0.08}$ & $4796^{+65}_{-30}$ & $3.069^{+0.009}_{-0.005}$ & $6.46^{+1.52}_{-1.35}$ & $1.17^{+0.07}_{-0.04}$ & $5.24^{+0.11}_{-0.10}$ & $4703^{+113}_{-140}$ & $3.071^{+0.004}_{-0.003}$ \\
7531119 & $5.70^{+1.02}_{-0.01}$ & $1.10^{+0.03}_{-0.02}$ & $7.62^{+0.08}_{-0.02}$ & $4659^{+62}_{-30}$ & $2.715^{+0.001}_{-0.003}$ & $5.55^{+1.64}_{-1.00}$ & $1.20^{+0.04}_{-0.05}$ & $7.95^{+0.17}_{-0.15}$ & $4620^{+138}_{-142}$ & $2.709^{+0.004}_{-0.004}$ \\
8057047 & $9.75^{+1.14}_{-1.24}$ & $1.02^{+0.02}_{-0.02}$ & $5.80^{+0.01}_{-0.08}$ & $4713^{+37}_{-62}$ & $2.918^{+0.003}_{-0.003}$ & $8.78^{+2.37}_{-1.87}$ & $1.08^{+0.03}_{-0.06}$ & $5.95^{+0.11}_{-0.13}$ & $4654^{+112}_{-149}$ & $2.918^{+0.003}_{-0.004}$ \\
8127582 & $7.42^{+1.06}_{-0.33}$ & $1.17^{+0.03}_{-0.04}$ & $5.36^{+0.03}_{-0.07}$ & $4752^{+39}_{-30}$ & $3.048^{+0.002}_{-0.005}$ & $6.70^{+1.45}_{-1.34}$ & $1.17^{+0.05}_{-0.04}$ & $5.42^{+0.10}_{-0.11}$ & $4662^{+132}_{-128}$ & $3.041^{+0.003}_{-0.004}$ \\
8552581 & $6.75^{+0.93}_{-1.21}$ & $1.15^{+0.05}_{-0.04}$ & $8.57^{+0.09}_{-0.13}$ & $4627^{+40}_{-45}$ & $2.632^{+0.003}_{-0.006}$ & $6.73^{+1.58}_{-1.55}$ & $1.15^{+0.05}_{-0.04}$ & $8.60^{+0.16}_{-0.18}$ & $4585^{+102}_{-144}$ & $2.628^{+0.004}_{-0.004}$ \\
8555265 & $7.71^{+4.14}_{-0.01}$ & $1.08^{+0.03}_{-0.18}$ & $8.76^{+0.10}_{-0.63}$ & $4644^{+80}_{-30}$ & $2.584^{+0.002}_{-0.015}$ & $5.52^{+1.30}_{-1.25}$ & $1.20^{+0.04}_{-0.05}$ & $9.14^{+0.20}_{-0.20}$ & $4620^{+107}_{-135}$ & $2.591^{+0.004}_{-0.004}$ \\
8613120 & $9.62^{+0.02}_{-1.79}$ & $1.06^{+0.04}_{-0.04}$ & $4.41^{+0.07}_{-0.06}$ & $4824^{+30}_{-30}$ & $3.173^{+0.006}_{-0.005}$ & $7.83^{+1.86}_{-1.45}$ & $1.08^{+0.05}_{-0.04}$ & $4.50^{+0.10}_{-0.08}$ & $4768^{+123}_{-159}$ & $3.168^{+0.004}_{-0.003}$ \\
8737047 & $8.61^{+1.32}_{-1.37}$ & $1.06^{+0.02}_{-0.06}$ & $12.36^{+0.12}_{-0.25}$ & $4467^{+33}_{-46}$ & $2.272^{+0.005}_{-0.005}$ & $7.54^{+1.97}_{-1.42}$ & $1.10^{+0.03}_{-0.04}$ & $12.74^{+0.26}_{-0.26}$ & $4386^{+98}_{-121}$ & $2.263^{+0.004}_{-0.004}$ \\
8782196 & $9.80^{+0.74}_{-1.78}$ & $1.00^{+0.02}_{-0.02}$ & $7.18^{+0.06}_{-0.06}$ & $4700^{+30}_{-45}$ & $2.723^{+0.003}_{-0.002}$ & $7.41^{+1.73}_{-1.59}$ & $1.08^{+0.05}_{-0.04}$ & $7.46^{+0.15}_{-0.13}$ & $4633^{+127}_{-147}$ & $2.726^{+0.004}_{-0.003}$ \\
8812397 & $7.97^{+1.18}_{-1.01}$ & $1.04^{+0.07}_{-0.04}$ & $7.21^{+0.15}_{-0.06}$ & $4672^{+47}_{-43}$ & $2.731^{+0.007}_{-0.005}$ & $5.87^{+1.37}_{-1.32}$ & $1.17^{+0.05}_{-0.04}$ & $7.63^{+0.16}_{-0.16}$ & $4623^{+138}_{-135}$ & $2.742^{+0.004}_{-0.004}$ \\
8872046 & $7.14^{+0.49}_{-0.88}$ & $1.15^{+0.02}_{-0.02}$ & $7.64^{+0.04}_{-0.05}$ & $4695^{+32}_{-44}$ & $2.731^{+0.003}_{-0.002}$ & $5.89^{+1.77}_{-1.23}$ & $1.17^{+0.03}_{-0.06}$ & $7.74^{+0.14}_{-0.17}$ & $4635^{+121}_{-142}$ & $2.724^{+0.003}_{-0.004}$ \\
8973228 & $6.07^{+0.85}_{-0.89}$ & $1.10^{+0.03}_{-0.02}$ & $8.68^{+0.07}_{-0.06}$ & $4668^{+38}_{-41}$ & $2.599^{+0.004}_{-0.003}$ & $6.52^{+1.72}_{-1.64}$ & $1.15^{+0.05}_{-0.05}$ & $8.78^{+0.18}_{-0.18}$ & $4602^{+97}_{-141}$ & $2.607^{+0.004}_{-0.004}$ \\
9009957 & $6.19^{+0.44}_{-1.11}$ & $1.17^{+0.02}_{-0.06}$ & $4.63^{+0.01}_{-0.10}$ & $4897^{+30}_{-38}$ & $3.175^{+0.001}_{-0.006}$ & $5.91^{+1.63}_{-1.34}$ & $1.15^{+0.05}_{-0.04}$ & $4.62^{+0.08}_{-0.09}$ & $4829^{+143}_{-144}$ & $3.168^{+0.003}_{-0.004}$ \\
9018909 & $7.40^{+0.53}_{-1.12}$ & $1.10^{+0.03}_{-0.02}$ & $7.13^{+0.07}_{-0.05}$ & $4693^{+46}_{-34}$ & $2.771^{+0.004}_{-0.003}$ & $6.71^{+1.74}_{-1.54}$ & $1.13^{+0.04}_{-0.05}$ & $7.26^{+0.14}_{-0.16}$ & $4653^{+126}_{-146}$ & $2.765^{+0.004}_{-0.004}$ \\
9025498 & $7.43^{+1.04}_{-0.90}$ & $1.11^{+0.04}_{-0.03}$ & $8.05^{+0.11}_{-0.08}$ & $4612^{+47}_{-35}$ & $2.670^{+0.003}_{-0.004}$ & $7.93^{+1.87}_{-1.97}$ & $1.11^{+0.04}_{-0.05}$ & $8.11^{+0.17}_{-0.16}$ & $4563^{+109}_{-138}$ & $2.663^{+0.003}_{-0.004}$ \\
9161451 & $7.38^{+0.76}_{-0.92}$ & $1.13^{+0.02}_{-0.05}$ & $5.57^{+0.04}_{-0.09}$ & $4749^{+53}_{-32}$ & $2.999^{+0.003}_{-0.006}$ & $6.22^{+1.46}_{-1.37}$ & $1.15^{+0.05}_{-0.04}$ & $5.67^{+0.11}_{-0.11}$ & $4707^{+112}_{-146}$ & $2.994^{+0.003}_{-0.004}$ \\
9205663 & $7.98^{+1.17}_{-1.01}$ & $1.04^{+0.02}_{-0.02}$ & $7.66^{+0.06}_{-0.06}$ & $4654^{+38}_{-47}$ & $2.684^{+0.003}_{-0.003}$ & $6.34^{+2.18}_{-1.18}$ & $1.13^{+0.07}_{-0.03}$ & $7.98^{+0.18}_{-0.17}$ & $4602^{+122}_{-145}$ & $2.689^{+0.004}_{-0.004}$ \\
9269772 & $7.18^{+1.01}_{-2.94}$ & $1.17^{+0.26}_{-0.09}$ & $8.22^{+1.24}_{-0.24}$ & $4607^{+30}_{-33}$ & $2.667^{+0.010}_{-0.029}$ & $7.18^{+1.98}_{-1.64}$ & $1.15^{+0.07}_{-0.04}$ & $8.24^{+0.16}_{-0.16}$ & $4546^{+89}_{-155}$ & $2.668^{+0.004}_{-0.004}$ \\
9277886 & $8.56^{+3.49}_{-2.41}$ & $1.08^{+0.05}_{-0.06}$ & $12.68^{+0.17}_{-0.24}$ & $4421^{+40}_{-47}$ & $2.259^{+0.011}_{-0.008}$ & $8.03^{+1.89}_{-1.72}$ & $1.11^{+0.04}_{-0.05}$ & $12.87^{+0.29}_{-0.30}$ & $4339^{+92}_{-128}$ & $2.258^{+0.004}_{-0.004}$ \\
9466520 & $9.16^{+0.01}_{-1.71}$ & $1.02^{+0.08}_{-0.02}$ & $8.33^{+0.20}_{-0.10}$ & $4633^{+30}_{-74}$ & $2.604^{+0.012}_{-0.001}$ & $7.24^{+1.61}_{-1.35}$ & $1.15^{+0.05}_{-0.04}$ & $8.69^{+0.18}_{-0.16}$ & $4552^{+82}_{-156}$ & $2.620^{+0.004}_{-0.003}$ \\
9515045 & $7.96^{+0.81}_{-0.99}$ & $1.11^{+0.02}_{-0.05}$ & $8.26^{+0.06}_{-0.14}$ & $4631^{+35}_{-41}$ & $2.647^{+0.002}_{-0.006}$ & $6.14^{+1.58}_{-1.49}$ & $1.17^{+0.07}_{-0.04}$ & $8.49^{+0.17}_{-0.19}$ & $4590^{+105}_{-137}$ & $2.650^{+0.004}_{-0.004}$ \\
9696576 & $5.19^{+0.72}_{-0.61}$ & $1.13^{+0.04}_{-0.02}$ & $11.27^{+0.16}_{-0.01}$ & $4582^{+48}_{-31}$ & $2.384^{+0.005}_{-0.001}$ & $5.63^{+1.85}_{-1.06}$ & $1.15^{+0.05}_{-0.04}$ & $11.49^{+0.21}_{-0.24}$ & $4477^{+129}_{-127}$ & $2.376^{+0.004}_{-0.004}$ \\
9716866 & $6.36^{+6.14}_{-0.98}$ & $1.17^{+0.05}_{-0.25}$ & $11.15^{+0.08}_{-2.46}$ & $4554^{+30}_{-59}$ & $2.418^{+0.101}_{-0.003}$ & $5.25^{+1.55}_{-1.12}$ & $1.20^{+0.04}_{-0.05}$ & $11.25^{+0.24}_{-0.22}$ & $4465^{+124}_{-123}$ & $2.410^{+0.004}_{-0.004}$ \\
9753097 & $8.48^{+0.59}_{-1.05}$ & $1.10^{+0.05}_{-0.02}$ & $7.20^{+0.11}_{-0.05}$ & $4661^{+44}_{-36}$ & $2.763^{+0.006}_{-0.002}$ & $5.94^{+1.77}_{-1.08}$ & $1.17^{+0.07}_{-0.04}$ & $7.44^{+0.16}_{-0.15}$ & $4612^{+115}_{-146}$ & $2.765^{+0.004}_{-0.004}$ \\
9773922 & $7.48^{+1.63}_{-0.49}$ & $1.04^{+0.02}_{-0.04}$ & $7.82^{+0.06}_{-0.10}$ & $4625^{+35}_{-30}$ & $2.665^{+0.003}_{-0.004}$ & $7.82^{+1.98}_{-1.76}$ & $1.08^{+0.05}_{-0.04}$ & $7.98^{+0.18}_{-0.16}$ & $4612^{+111}_{-144}$ & $2.664^{+0.004}_{-0.004}$ \\
10023439 & $9.10^{+1.14}_{-0.62}$ & $1.02^{+0.08}_{-0.04}$ & $7.39^{+0.20}_{-0.11}$ & $4617^{+34}_{-30}$ & $2.707^{+0.009}_{-0.004}$ & $6.93^{+1.88}_{-1.42}$ & $1.15^{+0.05}_{-0.05}$ & $7.78^{+0.14}_{-0.17}$ & $4585^{+109}_{-144}$ & $2.714^{+0.004}_{-0.004}$ \\
10087394 & $6.92^{+0.47}_{-1.01}$ & $1.13^{+0.04}_{-0.05}$ & $8.07^{+0.13}_{-0.13}$ & $4643^{+38}_{-48}$ & $2.675^{+0.005}_{-0.007}$ & $6.72^{+1.49}_{-1.82}$ & $1.15^{+0.05}_{-0.04}$ & $8.19^{+0.17}_{-0.15}$ & $4596^{+89}_{-149}$ & $2.673^{+0.004}_{-0.004}$ \\
10152909 & $9.74^{+1.48}_{-4.59}$ & $1.06^{+0.14}_{-0.06}$ & $6.79^{+0.50}_{-0.13}$ & $4613^{+58}_{-30}$ & $2.794^{+0.006}_{-0.006}$ & $7.43^{+1.68}_{-1.58}$ & $1.10^{+0.05}_{-0.04}$ & $6.93^{+0.14}_{-0.14}$ & $4627^{+118}_{-139}$ & $2.797^{+0.004}_{-0.004}$ \\
10157507 & $8.49^{+3.35}_{-2.40}$ & $1.04^{+0.04}_{-0.14}$ & $7.34^{+0.53}_{-0.09}$ & $4718^{+51}_{-56}$ & $2.724^{+0.008}_{-0.056}$ & $5.30^{+1.44}_{-1.00}$ & $1.20^{+0.04}_{-0.05}$ & $7.71^{+0.16}_{-0.15}$ & $4663^{+123}_{-153}$ & $2.736^{+0.003}_{-0.004}$ \\
10219075 & $9.08^{+1.31}_{-1.19}$ & $1.06^{+0.02}_{-0.04}$ & $5.12^{+0.04}_{-0.07}$ & $4728^{+61}_{-33}$ & $3.043^{+0.003}_{-0.005}$ & $7.92^{+1.91}_{-1.87}$ & $1.08^{+0.03}_{-0.04}$ & $5.20^{+0.09}_{-0.10}$ & $4740^{+126}_{-133}$ & $3.034^{+0.004}_{-0.003}$ \\
10223027 & $8.50^{+1.58}_{-4.70}$ & $1.13^{+0.13}_{-0.03}$ & $8.17^{+1.08}_{-0.08}$ & $4581^{+48}_{-30}$ & $2.661^{+0.004}_{-0.060}$ & $10.43^{+2.47}_{-2.49}$ & $1.06^{+0.04}_{-0.04}$ & $8.04^{+0.16}_{-0.17}$ & $4504^{+115}_{-136}$ & $2.649^{+0.004}_{-0.004}$ \\
10547370 & $8.82^{+1.87}_{-0.81}$ & $1.04^{+0.09}_{-0.04}$ & $9.57^{+0.34}_{-0.13}$ & $4521^{+46}_{-32}$ & $2.488^{+0.007}_{-0.004}$ & $6.78^{+1.76}_{-1.23}$ & $1.13^{+0.04}_{-0.05}$ & $9.94^{+0.20}_{-0.19}$ & $4477^{+105}_{-145}$ & $2.492^{+0.003}_{-0.004}$ \\
10548741 & $7.66^{+1.47}_{-1.40}$ & $1.08^{+0.03}_{-0.04}$ & $7.25^{+0.07}_{-0.98}$ & $4661^{+66}_{-30}$ & $2.758^{+0.101}_{-0.009}$ & $5.87^{+1.36}_{-1.29}$ & $1.20^{+0.04}_{-0.05}$ & $7.58^{+0.15}_{-0.15}$ & $4611^{+120}_{-139}$ & $2.753^{+0.004}_{-0.004}$ \\
10549732 & $6.34^{+1.18}_{-0.06}$ & $1.20^{+0.02}_{-0.03}$ & $8.97^{+0.07}_{-0.04}$ & $4605^{+30}_{-36}$ & $2.607^{+0.003}_{-0.004}$ & $7.73^{+1.51}_{-1.39}$ & $1.11^{+0.06}_{-0.03}$ & $8.89^{+0.17}_{-0.15}$ & $4487^{+128}_{-138}$ & $2.587^{+0.004}_{-0.004}$ \\
10855512 & $7.64^{+1.12}_{-0.68}$ & $1.11^{+0.02}_{-0.05}$ & $7.64^{+0.10}_{-0.13}$ & $4684^{+34}_{-43}$ & $2.716^{+0.002}_{-0.006}$ & $5.91^{+1.76}_{-1.07}$ & $1.17^{+0.05}_{-0.06}$ & $7.83^{+0.14}_{-0.18}$ & $4615^{+126}_{-137}$ & $2.715^{+0.004}_{-0.004}$ \\
10960798 & $8.21^{+1.26}_{-1.07}$ & $1.08^{+0.03}_{-0.02}$ & $7.71^{+0.08}_{-0.07}$ & $4613^{+39}_{-46}$ & $2.696^{+0.003}_{-0.003}$ & $7.09^{+1.80}_{-1.56}$ & $1.15^{+0.05}_{-0.04}$ & $7.97^{+0.15}_{-0.17}$ & $4555^{+106}_{-147}$ & $2.695^{+0.004}_{-0.003}$ \\
10989954 & $7.19^{+0.48}_{-1.09}$ & $1.10^{+0.03}_{-0.02}$ & $8.10^{+0.09}_{-0.04}$ & $4639^{+36}_{-44}$ & $2.660^{+0.003}_{-0.003}$ & $6.72^{+1.52}_{-1.52}$ & $1.17^{+0.05}_{-0.04}$ & $8.38^{+0.15}_{-0.20}$ & $4563^{+105}_{-141}$ & $2.659^{+0.004}_{-0.004}$ \\
10992345 & $8.52^{+0.92}_{-0.88}$ & $1.10^{+0.02}_{-0.06}$ & $8.10^{+0.04}_{-0.17}$ & $4624^{+54}_{-43}$ & $2.659^{+0.001}_{-0.008}$ & $7.21^{+1.92}_{-1.65}$ & $1.13^{+0.04}_{-0.05}$ & $8.20^{+0.17}_{-0.15}$ & $4599^{+109}_{-135}$ & $2.658^{+0.004}_{-0.004}$ \\
11027337 & $8.60^{+1.20}_{-1.08}$ & $1.04^{+0.04}_{-0.04}$ & $10.37^{+0.18}_{-0.14}$ & $4538^{+43}_{-37}$ & $2.417^{+0.003}_{-0.006}$ & $5.98^{+1.78}_{-1.12}$ & $1.15^{+0.05}_{-0.04}$ & $10.85^{+0.21}_{-0.23}$ & $4446^{+139}_{-91}$ & $2.424^{+0.004}_{-0.004}$ \\
11099193 & $7.18^{+1.02}_{-0.47}$ & $1.17^{+0.03}_{-0.02}$ & $8.55^{+0.06}_{-0.12}$ & $4615^{+69}_{-35}$ & $2.640^{+0.005}_{-0.002}$ & $7.20^{+1.61}_{-1.68}$ & $1.15^{+0.05}_{-0.04}$ & $8.55^{+0.17}_{-0.17}$ & $4546^{+106}_{-143}$ & $2.634^{+0.004}_{-0.004}$ \\
11293804 & $7.46^{+1.08}_{-1.34}$ & $1.06^{+0.07}_{-0.02}$ & $10.30^{+0.27}_{-0.07}$ & $4520^{+50}_{-30}$ & $2.436^{+0.004}_{-0.003}$ & $5.61^{+1.70}_{-1.02}$ & $1.17^{+0.05}_{-0.04}$ & $10.89^{+0.21}_{-0.21}$ & $4468^{+115}_{-132}$ & $2.433^{+0.003}_{-0.004}$ \\
11557992 & $6.73^{+0.51}_{-0.85}$ & $1.13^{+0.09}_{-0.03}$ & $8.17^{+0.20}_{-0.07}$ & $4622^{+62}_{-38}$ & $2.665^{+0.008}_{-0.005}$ & $6.44^{+1.80}_{-1.33}$ & $1.17^{+0.07}_{-0.04}$ & $8.35^{+0.16}_{-0.17}$ & $4569^{+94}_{-133}$ & $2.665^{+0.004}_{-0.004}$ \\
11607019 & $10.77^{+1.50}_{-1.46}$ & $0.96^{+0.04}_{-0.02}$ & $8.27^{+0.14}_{-0.08}$ & $4645^{+30}_{-43}$ & $2.580^{+0.004}_{-0.003}$ & $8.51^{+1.98}_{-2.03}$ & $1.08^{+0.05}_{-0.04}$ & $8.76^{+0.16}_{-0.17}$ & $4530^{+134}_{-162}$ & $2.583^{+0.004}_{-0.004}$ \\
11671621 & $7.75^{+0.48}_{-4.53}$ & $1.17^{+0.15}_{-0.07}$ & $8.51^{+1.11}_{-0.23}$ & $4588^{+48}_{-40}$ & $2.642^{+0.003}_{-0.049}$ & $7.27^{+2.01}_{-1.40}$ & $1.15^{+0.05}_{-0.04}$ & $8.49^{+0.17}_{-0.17}$ & $4517^{+113}_{-144}$ & $2.639^{+0.004}_{-0.003}$ \\
11759220 & $8.26^{+7.83}_{-1.01}$ & $1.17^{+0.03}_{-0.19}$ & $12.36^{+0.07}_{-0.81}$ & $4424^{+43}_{-37}$ & $2.318^{+0.003}_{-0.020}$ & $6.39^{+1.76}_{-1.17}$ & $1.20^{+0.04}_{-0.07}$ & $12.59^{+0.23}_{-0.27}$ & $4347^{+114}_{-122}$ & $2.308^{+0.004}_{-0.004}$ \\
11811203 & $6.08^{+0.87}_{-0.73}$ & $1.10^{+0.05}_{-0.02}$ & $9.63^{+0.17}_{-0.08}$ & $4610^{+40}_{-41}$ & $2.511^{+0.008}_{-0.004}$ & $6.33^{+1.15}_{-1.15}$ & $1.13^{+0.07}_{-0.03}$ & $9.79^{+0.20}_{-0.18}$ & $4534^{+147}_{-152}$ & $2.510^{+0.004}_{-0.004}$ \\
11875246 & $8.19^{+0.32}_{-1.05}$ & $1.10^{+0.07}_{-0.04}$ & $6.37^{+0.15}_{-0.09}$ & $4681^{+36}_{-35}$ & $2.870^{+0.008}_{-0.005}$ & $8.83^{+2.15}_{-1.85}$ & $1.08^{+0.05}_{-0.04}$ & $6.37^{+0.13}_{-0.13}$ & $4590^{+116}_{-147}$ & $2.865^{+0.003}_{-0.004}$ \\
11972774 & $5.21^{+0.39}_{-0.65}$ & $1.24^{+0.02}_{-0.04}$ & $8.88^{+0.07}_{-0.11}$ & $4665^{+39}_{-43}$ & $2.630^{+0.002}_{-0.004}$ & $5.59^{+1.60}_{-1.04}$ & $1.20^{+0.04}_{-0.07}$ & $8.79^{+0.19}_{-0.21}$ & $4601^{+117}_{-144}$ & $2.622^{+0.004}_{-0.004}$ \\
12116393 & $9.08^{+0.33}_{-1.64}$ & $1.06^{+0.05}_{-0.04}$ & $7.66^{+0.12}_{-0.11}$ & $4642^{+45}_{-43}$ & $2.693^{+0.004}_{-0.006}$ & $7.25^{+1.97}_{-1.71}$ & $1.11^{+0.06}_{-0.03}$ & $7.86^{+0.17}_{-0.16}$ & $4584^{+123}_{-143}$ & $2.693^{+0.004}_{-0.004}$ \\
12884116 & $12.27^{+0.75}_{-1.42}$ & $0.96^{+0.04}_{-0.04}$ & $8.14^{+0.12}_{-0.13}$ & $4639^{+40}_{-35}$ & $2.595^{+0.005}_{-0.005}$ & $7.42^{+1.78}_{-1.74}$ & $1.11^{+0.04}_{-0.05}$ & $8.63^{+0.19}_{-0.16}$ & $4596^{+106}_{-144}$ & $2.608^{+0.004}_{-0.004}$ \\
\hline
\end{longtable}}
\label{app:results}
\end{document}